%% file: Paper19.tex
\documentclass[submission,copyright,creativecommons]{eptcs}
\providecommand{\event}{TTC 2011} 
\usepackage{latexsym}
\usepackage{amssymb}
\usepackage{extarrows}
\usepackage{epsfig}
\usepackage{graphicx}
\usepackage{amsmath}
\usepackage{fancyhdr}
\usepackage{cite}
\usepackage{multirow}
\usepackage{listings}
\usepackage{booktabs}
\usepackage{amsfonts}
\usepackage{times}
\usepackage{fleqn}
\usepackage{macros}
\usepackage{algorithm}
\usepackage{algorithmic}
\usepackage{hyperref}
\usepackage{breakurl}

\newcommand{\hide}[1]{}

\title{\centering Solving the TTC 2011 Compiler Optimization Case with QVTR-XSLT}
\author{Dan Li \thanks{On leave from Guizhou Academy of Sciences, Guizhou, China}, Xiaoshan Li
\institute{Faculty of Science and Technology, University of Macau, China}
\email{lidan@iist.unu.edu, xsl@umac.mo}
\and
Volker Stolz
\institute{Department of Informatics, University of Oslo, Norway \& UNU-IIST, Macau, China}
\email{stolz@ifi.uio.no}
}

\def\titlerunning{Solving the TTC 2011 Compiler Optimization Case with QVTR-XSLT}
\def\authorrunning{Dan Li, Xiaoshan Li \& Volker Stolz}

\begin{document}

\maketitle

\section{Introduction \label{se:introduction}}

In this short paper we present our solution for the Compiler Optimization case study \cite{compileroptimizationcase} of the Transformation Tool Contest (TTC) 2011 using the \href{http://rcos.iist.unu.edu/qvttoxslt/}{QVTR-XSLT tool} \cite{LiQVTUMLFM2011}.
The tool supports editing and execution of the graphical notation of QVT Relations language \cite{OMG_QVT20}.

The case study addresses the problem of optimizing the \emph{intermediate representation} of compiled program code.
This problem consists of two tasks: local optimization and instruction selection.
The first task mainly concerns \emph{constant folding} which evaluates operations with only constant operands, corresponding control flows are also optimized.
The instruction selection task transforms the intermediate representation into a target representation of similar structure.
The SHARE demo related to the paper can be found at \cite{QVTRXSLTsolutions}.

We begin by giving a brief introduction of the QVTR-XSLT tool in Section~\ref{se:qvtrtool}.
Section~\ref{se:casestudy} explains the design of transformations for the case study.
We discuss the experimental result and evaluation of the solution against the criteria given in the case specification in Section~\ref{se:evaluation}.

\input{sec_QVTXSLT-2}

\section{Transformation Design \label{se:casestudy}}

\paragraph{The metamodel.}

\begin{figure}[!h]
\begin{center}
   \includegraphics[width=0.4\linewidth]{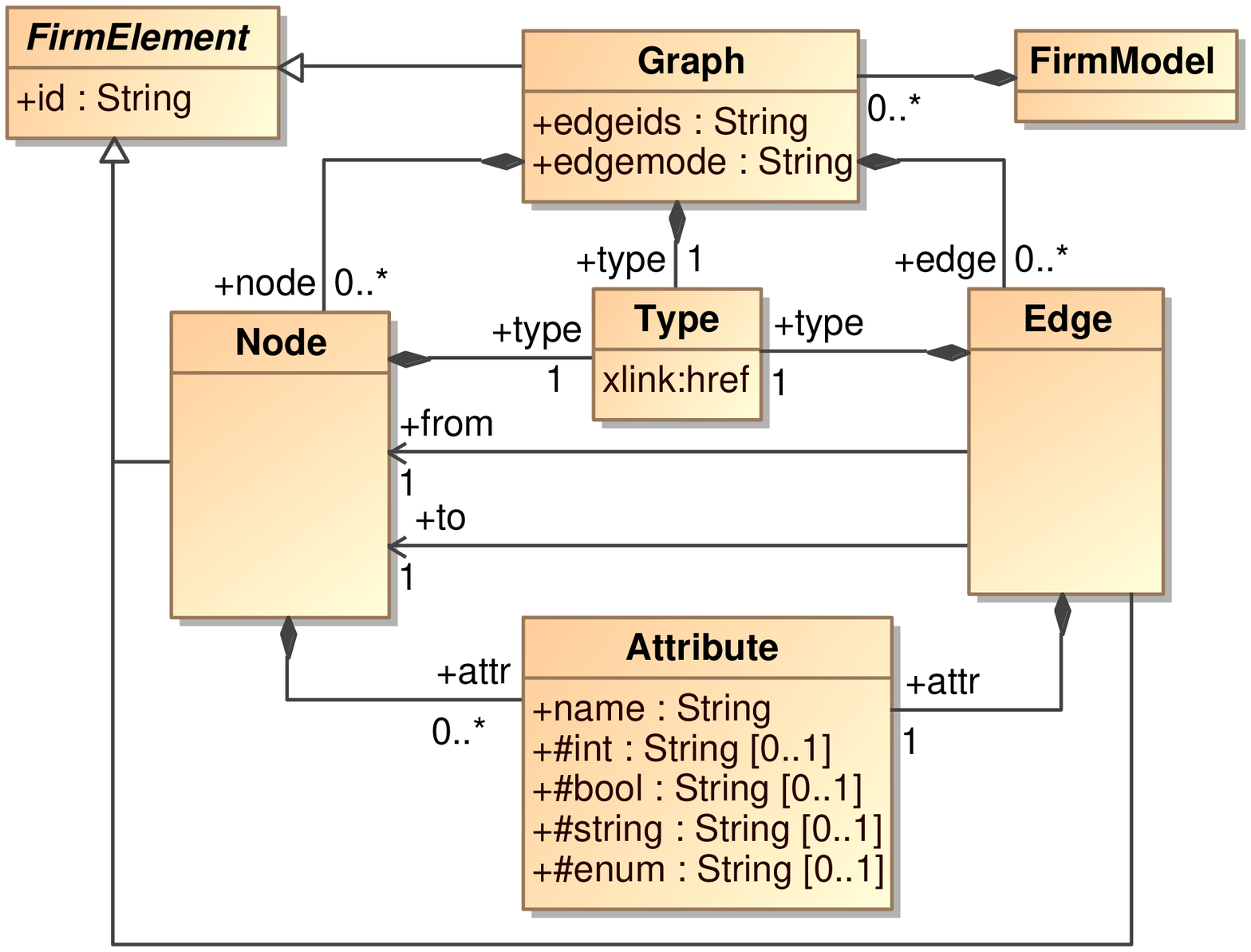}
\end{center}
\vspace*{-1.0\baselineskip}
   \caption{The metamodel \hide{\emph{Intermediate}} for IR}
\label{fig:Intermediate}
\end{figure}

As the first step of the transformation design, we define a simple metamodel for the intermediate representation (IR) of the \textsc{Firm} model, as shown in Fig.~\ref{fig:Intermediate}. In the metamodel, a \textsc{Firm} model consists of \emph{Graphs}, and the transformations only deal with graphs of type \emph{Default Graph}.
Within a graph, a \emph{Node} represents an operation, and the type of the operation is decided by the \emph{xlink:href} property of its \emph{Type} element. The property may be \emph{\#Jmp}, \emph{\#Add}, \emph{\#Mul}, or \emph{\#And}, etc.
It also could be \emph{\#StartBlock}, \emph{\#Block}, or \emph{\#End} for a control flow node.
A node may also own some \emph{Attributes}, each of which has a name, and some values of different types.
\emph{Edges} specify the dependencies between nodes.
An edge also has a type, such as \emph{\#Dataflow} or \emph{\#ControlFlow}, and a \emph{position}.

We define a set of well-formedness rules as OCL class invariants for the metamodel:

\noindent $\bullet$ \ An instance of class \emph{Attribute} has only two properties (one of them is the \emph{name}).

\begin{lstlisting}[language={},numbers=none,basicstyle=\footnotesize,%
morekeywords={context, inv, self}, keywordstyle=\bfseries]
context Attribute  inv twoProperties: self.getAllProperties() %$ \rightarrow $% size()=2
\end{lstlisting}

\noindent where the \emph{getAllProperties}() returns all properties of an instance.

\noindent $\bullet$ \ A \emph{FirmMode} has at least one default graph.

\begin{lstlisting}[language={},numbers=none,basicstyle=\footnotesize,%
morekeywords={context, inv, self}, keywordstyle=\bfseries]
context Graph inv hasDefault: Graph.allInstances()%$ \rightarrow $%select(id='DefaultGraph')%$ \rightarrow $%size()>=1
\end{lstlisting}

\noindent $\bullet$ \  Within a default graph, the \emph{id} attribute is the unique identifier.
\begin{lstlisting}[language={},numbers=none,basicstyle=\footnotesize,%
morekeywords={context, inv, self}, keywordstyle=\bfseries]
context Graph  inv uniqueId:
   Graph.allInstances()%$ \rightarrow $%select(id='DefaultGraph')%$ \rightarrow $% forAll(g|g.node.id%$ \rightarrow $%asBag()
        %$ \rightarrow $%union(g.edge.id%$ \rightarrow $%asBag())%$ \rightarrow $%isUnique(id))
\end{lstlisting}

\paragraph{Local optimizations.}

This task is a typical \emph{in-place} transformation, in which both the input and output models are the same, and the execution of one rule could affect its subsequent rules.
Since the execution unit of QVTR is a transformation, the optimization task is actually accomplished by a chain of executions of the transformation; each execution makes some changes to the model, and its output serves as the input of the following execution.
Execution will stop if no more changes happen. This process is automated if running in the transformation runner.

The complete transformation consists of 13 relations, 9 queries and 2 functions (see Appendix \ref{ap:localopt}).
Some of the relations are used for auxiliary purposes, such as removing \emph{blocks}, \emph{nodes} and \emph{edges} from the model, or changing the \emph{position} of an edge (Fig.~\ref{fig:RemoveEdge}--\ref{fig:RemoveNode}).
As in the model there is no direct navigation between nodes, or from a node to its connected edges, queries are defined to retrieve information such as the incoming and outgoing edges of a node, or a node's original and destination nodes.  (Fig.~\ref{fig:GetEdgeInNode}--\ref{fig:GetToNode}).
Some queries are functionally overlapping, as we want to use appropriate query names in different situations.
All mathematical and logical calculations are performed by two functions.  (Fig~\ref{fig:CalcuLogic}--\ref{fig:CalcuMatch}) .
Because of the limited mathematical abilities of OCL expressions, we only deal with mathematical operations of \emph{Add}, \emph{Sub}, \emph{Mul} and \emph{Div}, and \emph{LESS}, \emph{EQUAL} and \emph{GREATER} for logic operations.

The main part of the transformation definition has 9 relations that will be discussed in the following:

\begin{itemize}
  \item \textbf{FirmModelTrans}: The transformation starts from this initial \emph{top level} relation, which matches a graph with type of \emph{DefaultGraph}, then the relation \emph{FoldOper} and relation group \emph{FoldNode} are invoked from the \emph{where} clause.
  \item \textbf{FoldOper}: The relation first checks whether the node is a binary operation of the four mathematical kinds or a \emph{Cmp}, then the query \emph{GetToData} is called to get its two operands. If two of them are \emph{Const}, the relation group \emph{DoFoldOper} is invoked.
  \item \textbf{DoFoldOper}: This group includes two relations: \emph{DoFoldCmp} and \emph{DoFoldMath}.
Inside the \emph{DoFoldCmp} relation, values of the two const operands of the \textsf{Cmp} node are compared with the \emph{CalcuLogic} function, and also compared with the incoming edges of the corresponding \textsf{Cond} node to decide which const operand should be removed. Then the \textsf{Cond} node and its connected edges are removed, and the \textsf{Cmp} node becomes a \textsf{Jmp} node.
The \emph{DoFoldMath} relation first calculates the mathematical result of the two consts, and then changes the operation node into a \textsf{Const} node and puts the result as the value of the node; finally the two outgoing edges of the operation node are removed.

  \item \textbf{FoldNode}: This group includes the following relations:
  \begin{itemize}
    \item \textbf{FoldPhi} and \textbf{DoFoldPhi}: If there is only one outgoing dataflow edge for a \textsf{Block} where a \textsf{Phi} is located in, the \emph{FoldPhi} relation invokes the \emph{DoFoldPhi} relation, and the latter removes the \textsf{Phi} node, relinks its users directly to the correct \textsf{Const}, and resets the \emph{position} of the linking edge.
    \item \textbf{FoldJmpBlock}:  The relation removes blocks which only contain a \textsf{Jmp} node.
    \item \textbf{FoldNoOutBlock}: Removes blocks without any outgoing control flow edges.
    \item \textbf{FoldIsolateConst}: Removes \textsf{Const} nodes which have no users.
  \end{itemize}

\end{itemize}

Two XSLT stylesheet are generated for the transformation, with a total of 480 lines.

\paragraph{Instruction selection.}

The transformation for instruction selection is designed as a source to target model transformation, while both source and target models share the same metamodel.
The complete transformation consists of 13 relations, 5 queries and 2 functions (see Appendix~\ref{ap:selecte}).
Many of the relations are used for trivial one-to-one copying from the source model to the target model (\emph{CopyAtt, CopyNode, EdgeToEdge, OtherGraph}).
The transformation starts from relation \emph{FirmMode}, and then the relation groups \emph{GraphToGraph} and \emph{NodeToNode} are sequentially invoked, the latter includes relation \emph{BinaryOp} and \emph{UniqueOp}.

The major work of the transformation is accomplished by the following relations:

\begin{itemize}
  \item \textbf{BinaryOp}: All binary operations are selected by the relation. For each operation \emph{op}, we change its type \emph{optp} to ``\emph{Target}"+\emph{optp},  and invoke relation \emph{MakeBinaryI}.
  \item \textbf{MakeBinaryI}: An additional new operation \emph{top} is created with type of ``\emph{Target}"+ \\ \emph{optp}+``\emph{I}", along with a new \emph{value} attribute, and all connected edges of \emph{op} are duplicated to \emph{top} using relation \emph{MakeEdge}. Moreover, if the commutative property of \emph{op} is \emph{false}, relation \emph{MakeNewConst} is invoked.
  \item \textbf{MakeNewConst}: The relation creates a new \textsf{Const} node in the start block, and an outgoing edge with \emph{position} 1 is also created to link \emph{top} to the const node.
  \item \textbf{UniqueOp}: All other operations marked with ``*" in the case specification are selected by the relation. The operation type \emph{optp} is changed to ``\emph{Target}"+\emph{optp}. If the operation is \textsf{Load} or \textsf{Store}, we invoke relation \emph{MakeLoadStoreI}.
  \item \textbf{MakeLoadStoreI}: A new \textsf{StoreI} or \textsf{LoadI} operation node is created, which has an additional \emph{symbol} attribute with a string value of ``global".
\end{itemize}

Similar to the task of local optimizations, queries are used to retrieve information from the model.
Function \emph{GetTargetName} computes the target type from the type of an operation, and \emph{GetNewId} generates a new identifier for a model element.
An XSLT stylesheet of 330 lines of code is generated for the transformation.


In addition, we implement the model validating rules given in the case specification as an independent transformation. The generated XSLT stylesheet for the transformation is about 280 lines of code. It outputs a HTML page showing the results.

\section{Experiments and Evaluation \label{se:evaluation}}

Using our transformation runner, we execute the transformations on the examples provided by the case in a laptop of Intel M330 2.13 GHz CPU, 3 GB memory, and running Windows 7 home version.
The DTD definition (second line) has to be removed from the .gxl file of each example to prevent the produce of additional namespace information.
The results are shown in Table~\ref{tab:performance}.
The execution time includes loading and saving model files from/to disk.

\begin{table}[t]
 \centering \caption{Result of the transformations for compiler optimization }
\label{tab:performance}
 \begin{tabular}{llll}
  \toprule
 Transformation & \ \ \ Example (.gxl) & \ \ \ Size (kb) & \ \ \ Time (ms) \\
  \midrule
   Local Optimizations   & \ \ \ min   & \ \ \ 36 & \ \ \ 155\\
                         & \ \ \ const  & \ \ \ 59 & \ \ \ 410\\
   Instruction Selection & \ \ \ const   &  \ \ \ 59 &  \ \ \ 15\\
                         & \ \ \ mem  & \ \ \ 198 & \ \ \ 850\\
                         & \ \ \ testcase  & \ \ \ 186 & \ \ \ 820\\
  \bottomrule
 \end{tabular}
\end{table}

Our solution has covered all examples of the two tasks of the case study, except \emph{zero.gxl}, which needs more math functions than our tool can provide, such as shifting and bit operations. 
As a high-level general purpose transformation languages, neither QVTR nor XSLT offer explicit parallelism, and leave this to a particular implementation.
We are not aware that any XSLT processor makes use of parallelism except for an Intel research prototype.

The performance and the memory needed are much dependent on the XSLT processor used, and we can see from the results our tool works well, as it completes in under one second for every example.
Our solution is \emph{pure}, since no other code (e.g. hand-crafted XSLT) is required for the transformation of the examples, except for the iterative runner which applies the transformation until the result stabilises.

\paragraph{Conclusion \label{se:conclusion} }

Based on the QVTR-XSLT tool, we define a transformation using the graphical notation of QVT Relations, and generate an XSLT program to execute the transformation.
Our contribution is two-fold:
we have provided a solution for the two tasks of the compiler optimization case study of TTC 2011, and shown that our QVT-XSLT engine translates those examples, so that they can be executed in a standard XSLT engine.

{\small
\noindent{\bf \\ Acknowledgements } Partially supported by the ARV and GAVES grants of the Macau Science and Technology Development Fund, and the Guizhou International Scientific Cooperation Project G[2011]7023 and GY[2010]3033.
}

\bibliographystyle{eptcs}
{\small \bibliography{bibtex/rcos,bibtex/managed}}

\newpage

\appendix
\input{TTC2011_Appendix_qvtrintro.tex}

\clearpage
\input{TTC2011_Appendix_opt.tex}

\clearpage
\input{TTC2011_Appendix_sel.tex}

\end{document}

%% file: sec_QVTXSLT-2.tex
\section{The QVTR-XSLT tool \label{se:qvtrtool}}

Model transformation is the core technology for model-driven development, and is used in software model refinement, evolution, refactoring and code generation.
To address the need for a common transformation language, the Object Management Group (OMG) proposed the Query/View/Transformation language (QVT) \cite{OMG_QVT20} standard.
QVT has a hybrid declarative/imperative nature.
In its declarative language, called QVT Relations (QVTR), a transformation is defined as a set of \emph{relations} (rules) between source and target models, each conforming to their respective metamodels.
Transformations are driven by a single, designated top-level relation.

QVTR combines a textual and a graphical notation.
In graphical syntax, a relation specifies how two object diagrams, called \emph{domain patterns}, relate to each other.
That is, the \emph{structural} matching of elements in the source- and target model is done diagrammatically.
Moreover, QVTR employs a textual language based on essential OCL \cite{OMG_OCL20} to define additional (non-structural) constrains in relations.
The graphical notation of QVTR provides a concise, intuitive and yet powerful way to specify transformations. However, currently there are very few tools supporting QVTR, and even fewer for its graphical notation.
\hide{
QVTR-XSLT is a tool supporting a \underline{slightly extended}\footnote{Hm, so you have some extensions, but you do not support ``full'' QVTR. This makes it very easy to attack the paper. What are the extensions?} version of the graphical notation of QVT Relations. It consists of two parts:
}

QVTR-XSLT supports the graphical notation of QVT Relations, and an execution engine for a subset of QVTR by means of XSLT programs.
It consists of two parts:

\begin{itemize}
  \item \textbf{Graphical Editor}: Building on top of \emph{MagicDraw UML} \cite{MagicDraw}, the editor has a graphical interface for defining metamodels as simple class diagrams, specifying QVTR relations and queries in graphical notation, validating the design, and saving the transformations as an XML file. 
  \item \textbf{Code generator}: It reads in the XML file, and generates an XSLT stylesheet for a transformation. 
\end{itemize}

\hide{
\vspace*{-1.0\baselineskip}

\begin{figure}[!h]
\begin{minipage}[t]{0.5\linewidth}
\vspace{0pt} \centering
\includegraphics[width=1.0\linewidth]{pics2/QVTRToolbar.eps}
\end{minipage}%
\begin{minipage}[t]{0.5\linewidth}
\vspace{0pt} \centering
\includegraphics[width=0.9\linewidth]{pics2/codegenerator.eps}
\end{minipage}%
\\[+2pt]
\begin{minipage}[c]{0.5\linewidth}
\caption{Toolbar of QVTR graphical editor}
\label{fig:QVTRToolbar}
\end{minipage}%
\begin{minipage}[c]{0.5\linewidth}
   \caption{XSLT code generator}
\label{fig:codegenerator}
\end{minipage}
\end{figure}
}

The outputs of the code generator are pure XSLT programs which can be directly executed under any XSLT processor on any platform.
Additionally, we have also developed a transformation runner, in the form of a Java program invoking the Saxon 9 XSLT processor, to facilitate the execution of generated XSLT stylesheets.

The QVTR-XSLT tool supports transformation parameters, transformation inheritance through rule overriding, and multiple input and output models.
Furthermore, \emph{in-place} transformations are defined as modifications (insert, remove, replace) of the existing model elements.
QVTR-XSLT-based transformations are used in the \href{http://rcos.iist.unu.edu}{rCOS Modeler} for use case-driven development of component- and object systems.


%% file: TTC2011_Appendix_qvtrintro.tex
\section{A Brief Introduction to QVT Relations \label{ap:qvtintro}}

QVT Relations (QVTR) is a declarative model transformation language proposed by the OMG as part of the MOF Query/View/Transformations (QVT) standard \cite{OMG_QVT20}.
QVTR specifies a \emph{transformation} as a set of \emph{relations} between
source and target metamodels.
A metamodel is defined in our tool as a simple class diagram.
In addition, a transformation may own some \emph{functions}, which are side-effect-free
operations, and \emph{queries} used to retrieve information from the models.

In the graphical notation, a \emph{relation} defines how two object diagrams, called \emph{domain
patterns}, relate to each other.
The object with  tag \emph{$\ll$domain$\gg$} is the \emph{root} of a domain pattern, and it also serves as a parameter of the relation.
In general, we assume the left domain pattern is the source domain, and the right the target domain.
An \emph{object} or a property of an object could be given a name that is taken as a \emph{variable}.
If the object is in the source domain pattern, then the object or the value of the property is bound to the variable.
Otherwise the object in target domain pattern means assigning the value of the variable to the object or property.
Note that a property variable in the diagrams may contain additional quote-characters that are an artefact of the visualization, and not string delimiters.

When a relation is executed, the source domain pattern is searched in the source model by way of
\emph{pattern matching} which starts from the domain root.
When a match is found, all variables defined in source domain pattern are bound to values.
The target domain pattern acts as a template to create corresponding objects and links in the target model using the values of the variables in the pattern.

A relation may define a pair of optional \emph{when}- and \emph{where}-clauses which consist of a set of OCL expressions.
The \emph{when}-clause indicates additional matching conditions for the relation.
And new variables can be defined in the \emph{where}-clause.
Other relations could be invoked in the \emph{where}-clause and variables can be passed as arguments.
A relation may also have \emph{primitive domains} in order to pass additional parameters between the relations.
Furthermore, a relation is either designed as a \emph{top-level} relation, or
a \emph{non-top-level} relation.
A \emph{top-level} relation is invoked from the transformation framework, and \emph{non-top-level} relations are invoked by other relations.

%% file: TTC2011_Appendix_opt.tex
\section{Transformation for Local Optimizations \label{ap:localopt}}

\noindent $\bullet$ \textbf{Transformation configuration:} name : \emph{TTC\_LocalOptimizations}, isInPlace : \emph{true},  rInPlace : \emph{true}, source : \emph{Intermediate}, sourceKey : \emph{id}, sourceName : \emph{original}, target: \emph{Intermediate}, targetKey:\emph{id}, targetName : \emph{optim}.

\vspace*{-0.5\baselineskip}

\subsection{QVTR relations}

\vspace*{-1.5\baselineskip}

\begin{figure}[!h]
\begin{minipage}[t]{0.5\linewidth}
\vspace{0pt} \centering
\includegraphics[width=.8\linewidth]{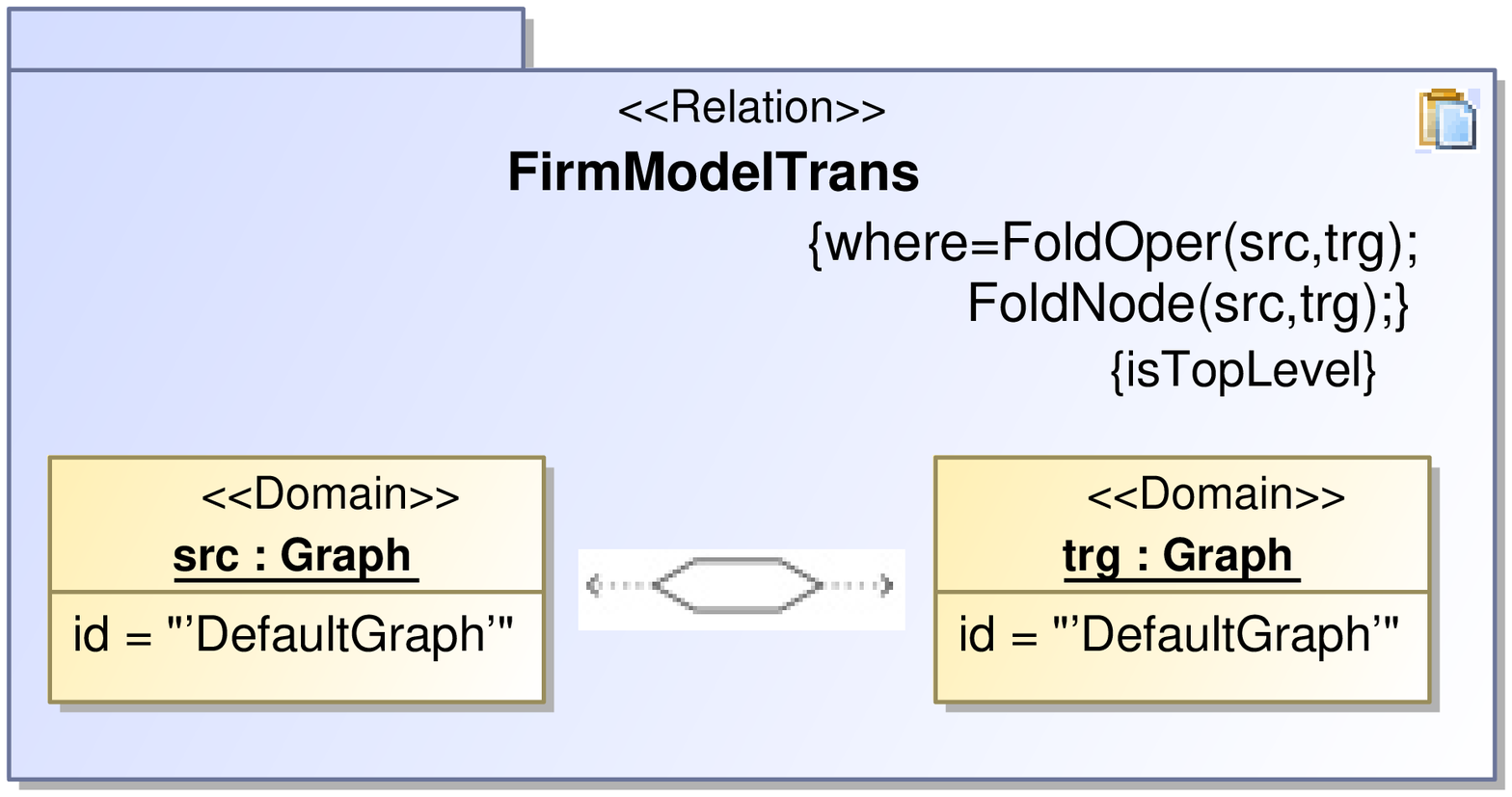}
\end{minipage}%
\begin{minipage}[t]{0.5\linewidth}
\vspace{0pt} \centering
\includegraphics[width=.8\linewidth]{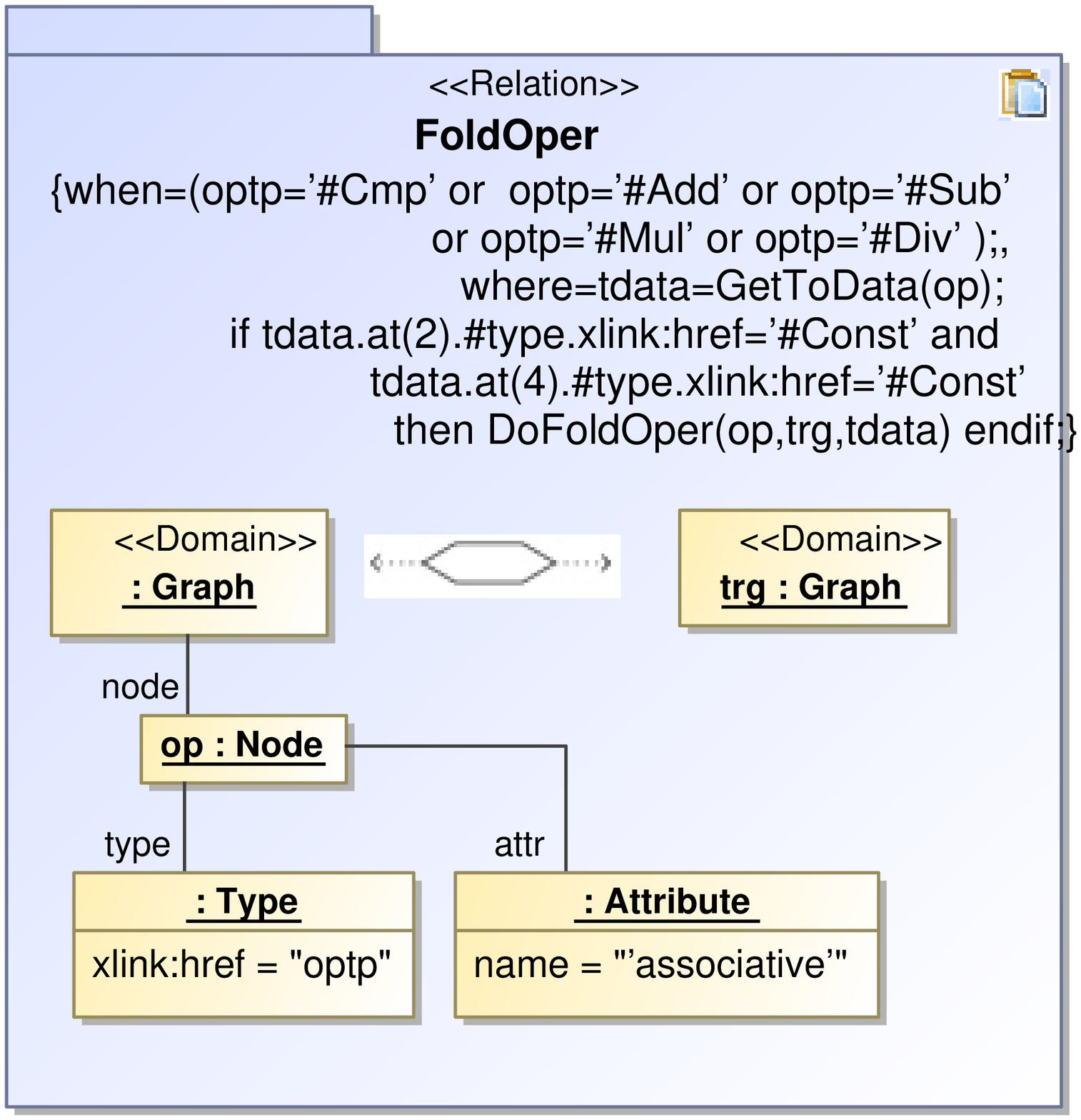}
\end{minipage}%
\\[+2pt]
\begin{minipage}[c]{0.5\linewidth}
\caption{Starting top level relation}
\label{fig:FirmModelTrans}
\end{minipage}%
\begin{minipage}[c]{0.5\linewidth}
   \caption{Select binary operation with two const operands}
\label{fig:FoldOper}
\end{minipage}
\end{figure}

\vspace*{-1.5\baselineskip}

\begin{figure}[!h]
\begin{center}
   \includegraphics[width=0.75\linewidth]{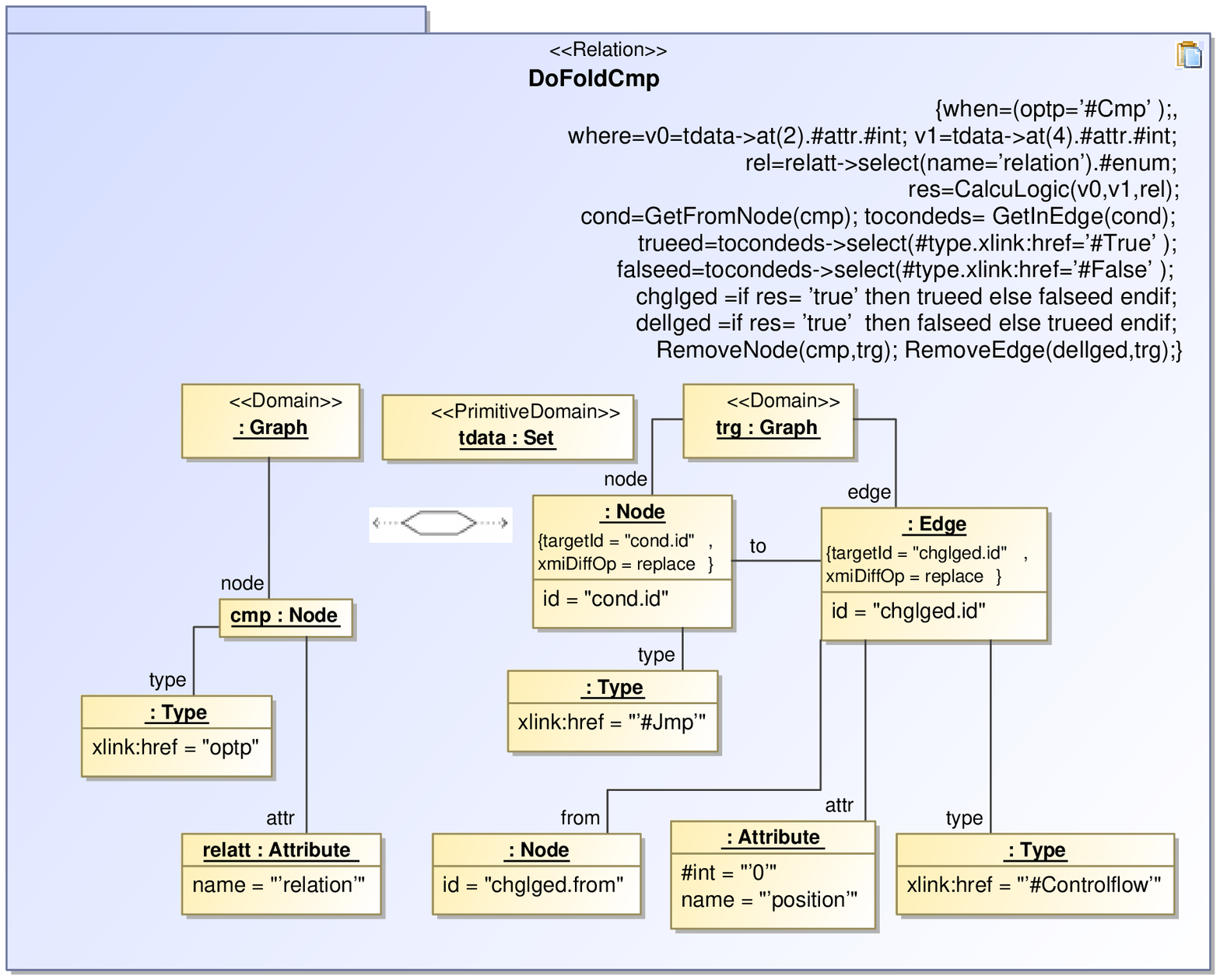}
\end{center}
\vspace*{-1.5\baselineskip}
   \caption{Cope with \textsf{Cmp} operation (relName: \emph{DoFoldOper}, rInPlace : \emph{true})}
\label{fig:DoFoldCmp}
\end{figure}


\begin{figure}[!h]
\begin{center}
   \includegraphics[width=0.65\linewidth]{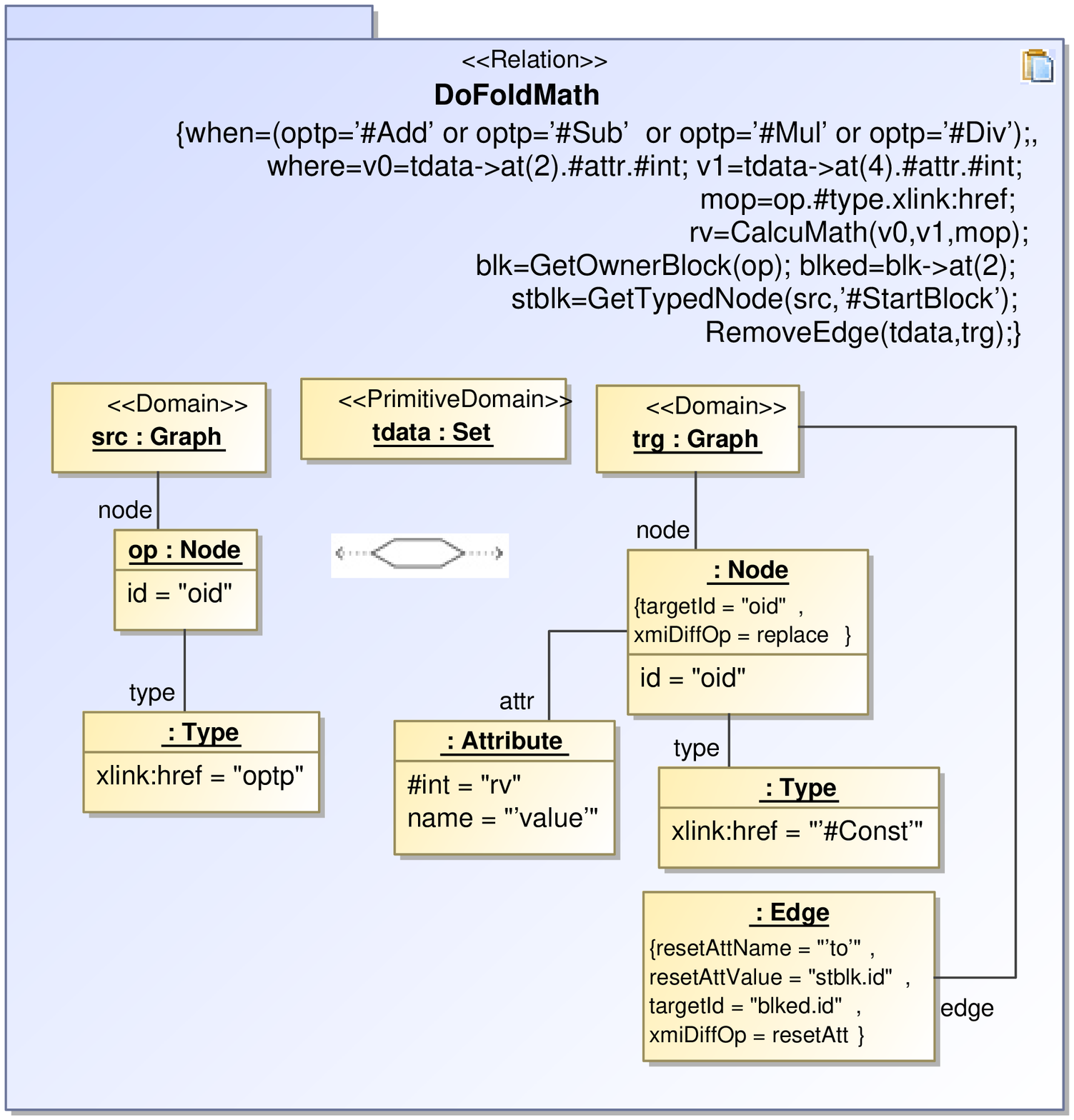}
\end{center}
\vspace*{-1.5\baselineskip}
   \caption{Cope with math operations (relName: \emph{DoFoldOper}, rInPlace : \emph{true})}
\label{fig:DoFoldMath}
\end{figure}

\vspace*{-1.5\baselineskip}

\begin{figure}[!h]
\begin{minipage}[t]{0.4\linewidth}
\vspace{0pt} \centering
\includegraphics[width=.9\linewidth]{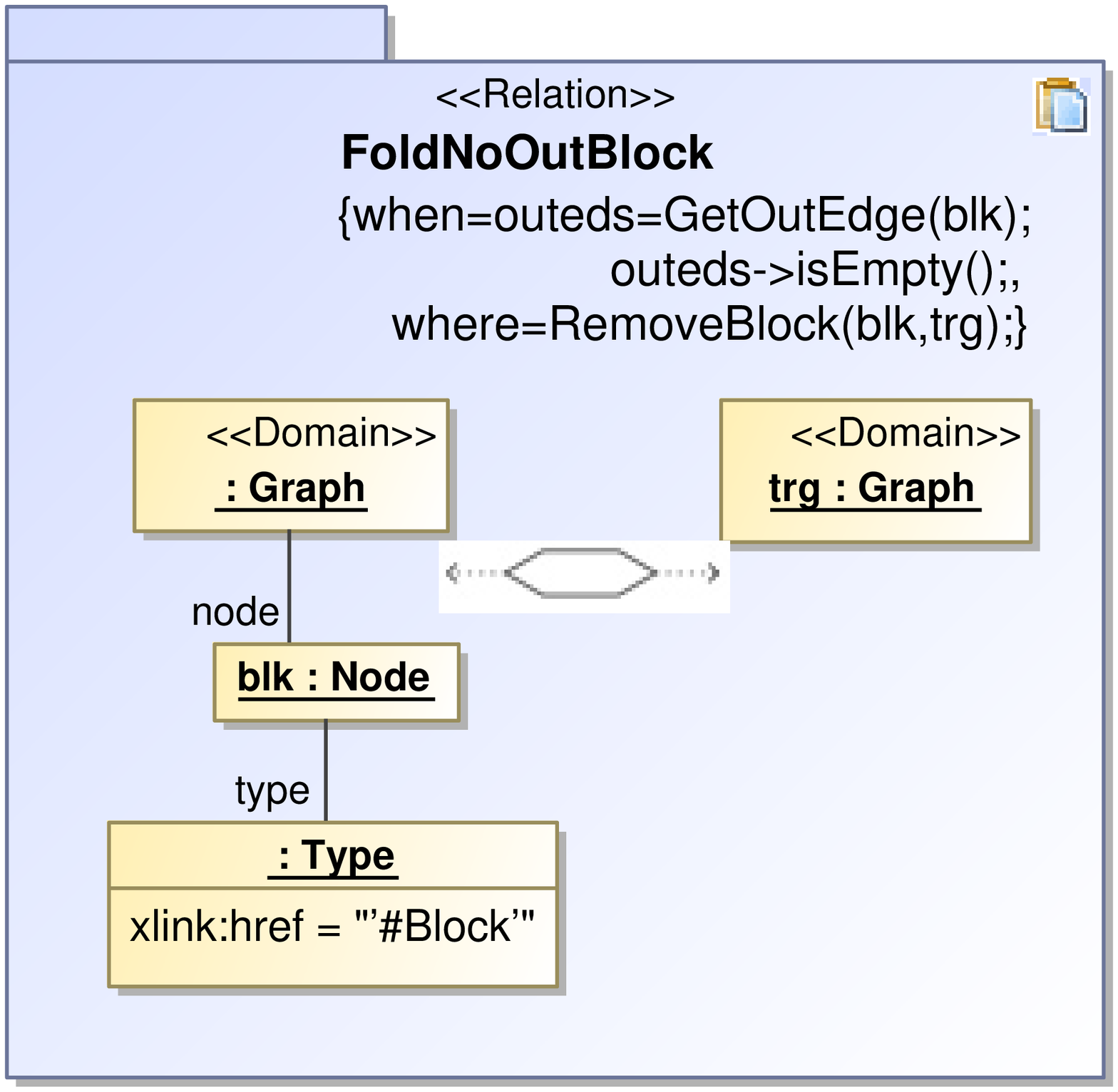}
\end{minipage}%
\begin{minipage}[t]{0.6\linewidth}
\vspace{0pt} \centering
\includegraphics[width=.9\linewidth]{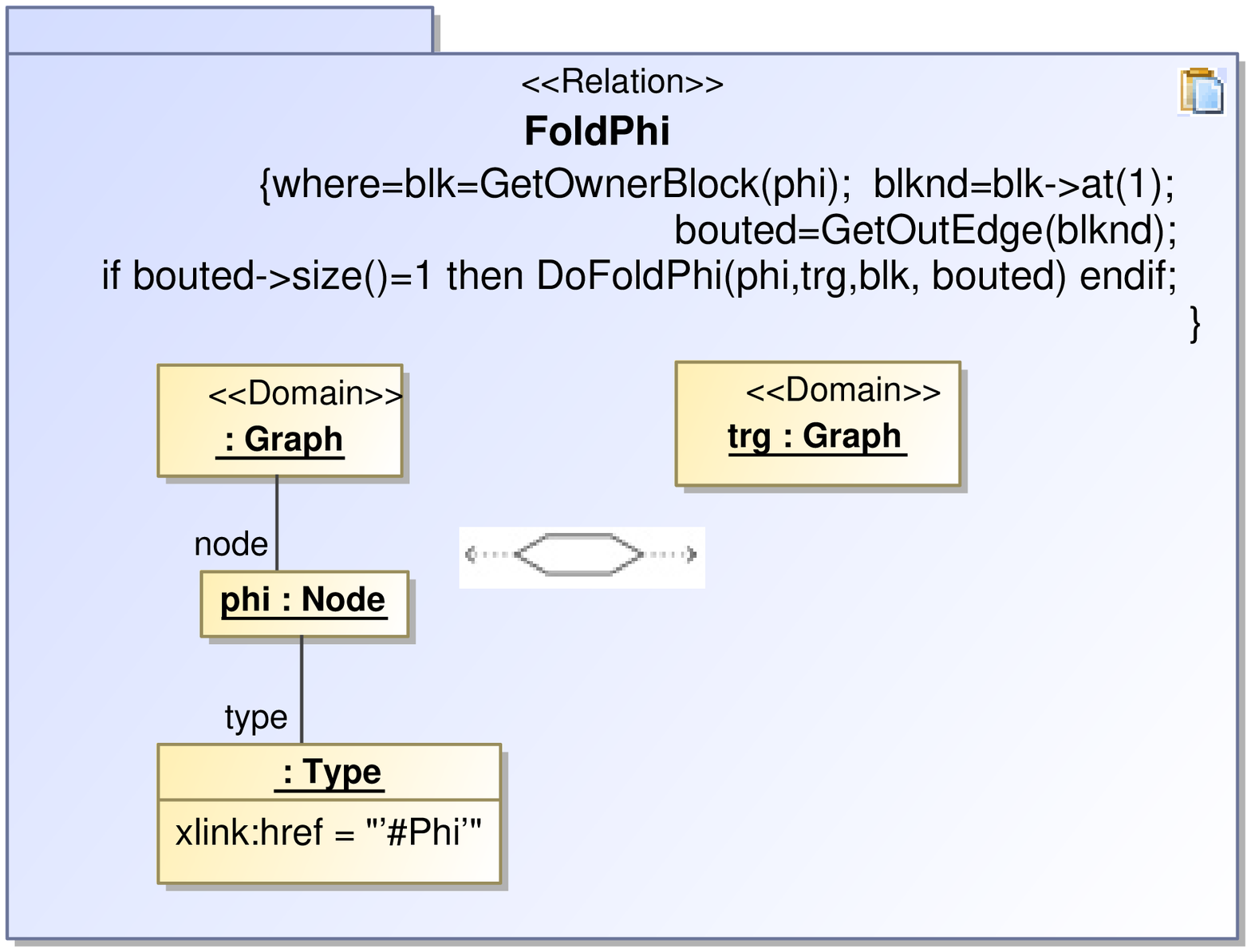}
\end{minipage}%
\\[+5pt]
\begin{minipage}[c]{0.4\linewidth}
\caption{Fold block without outgoing edge (relName: \emph{FoldNode}, rInPlace : \emph{true})}
\label{fig:FoldNoOutBlock}
\end{minipage}%
\hspace{6pt}
\begin{minipage}[c]{0.5\linewidth}
   \caption{Select a \textsf{Phi} owned by a block with only out control edge (relName: \emph{FoldNode})}
\label{fig:FoldPhi}
\end{minipage}
\end{figure}

\begin{figure}[!h]
\begin{center}
   \includegraphics[width=0.8\linewidth]{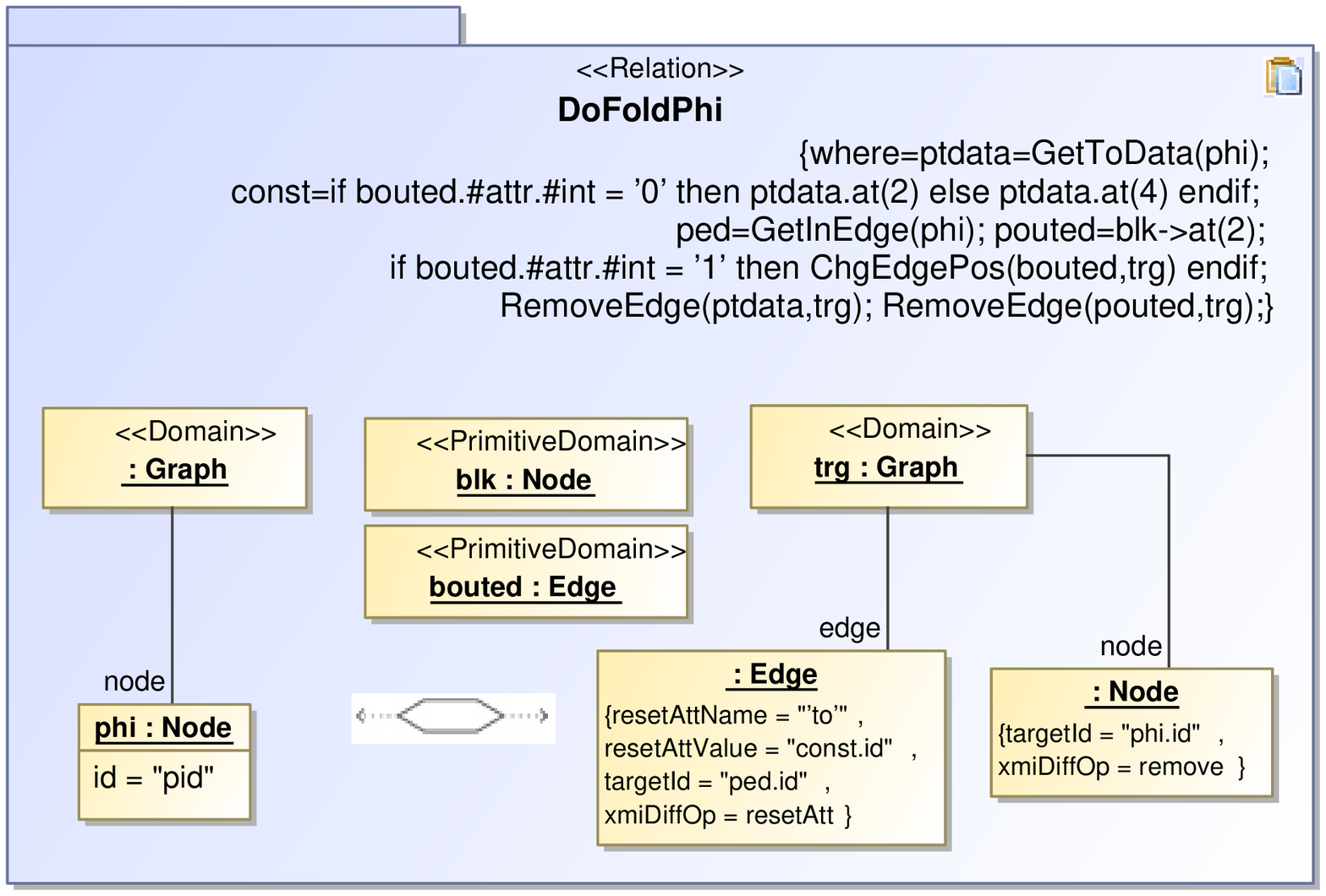}
\end{center}
\vspace*{-1.5\baselineskip}
   \caption{Fold a \textsf{Phi} node (rInPlace : \emph{true})}
\label{fig:DoFoldPhi}
\end{figure}

\begin{figure}[!h]
\begin{minipage}[t]{0.4\linewidth}
\vspace{0pt} \centering
\includegraphics[width=1.0\linewidth]{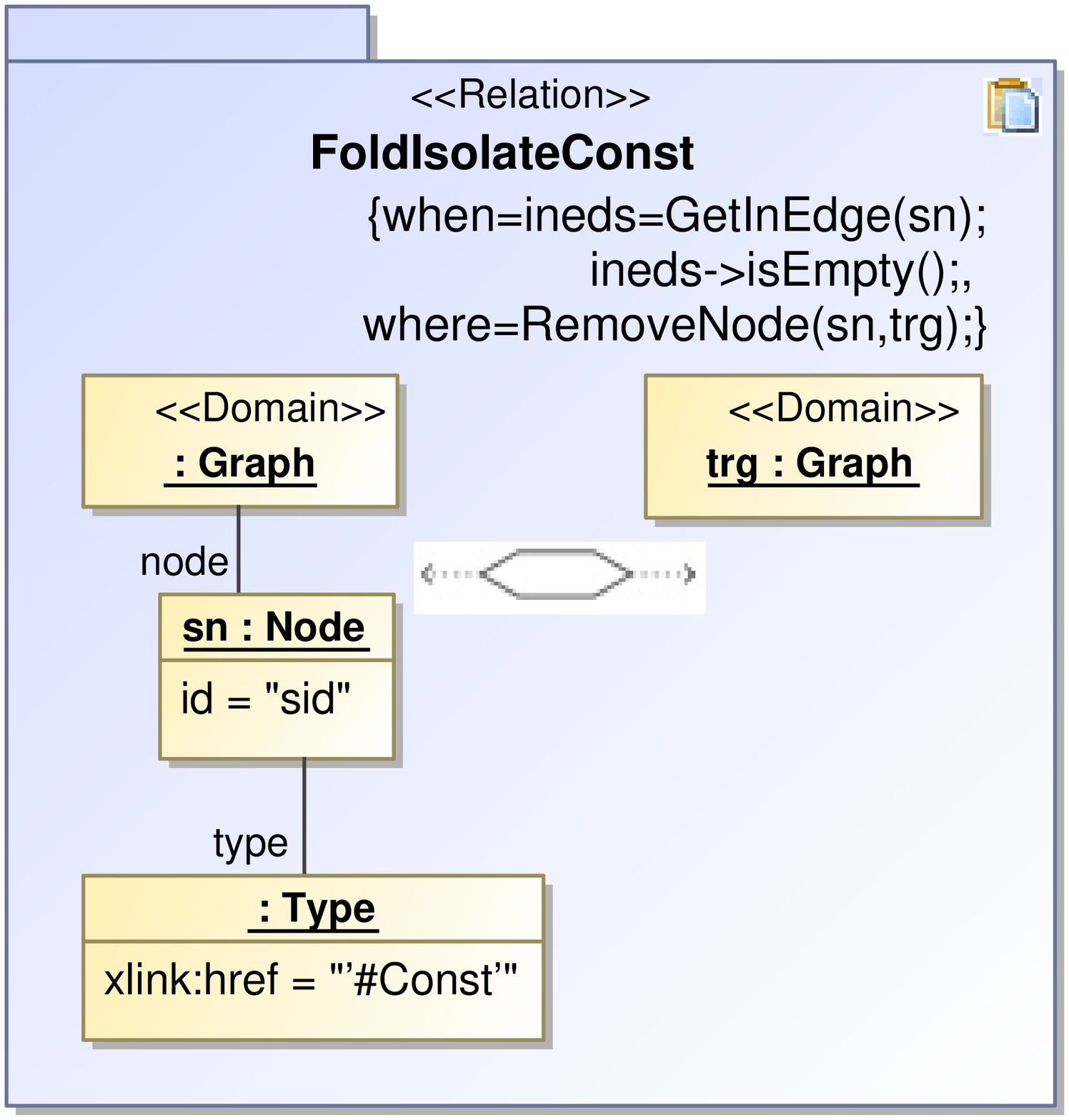}
\end{minipage}%
\begin{minipage}[t]{0.6\linewidth}
\vspace{0pt} \centering
\includegraphics[width=1.0\linewidth]{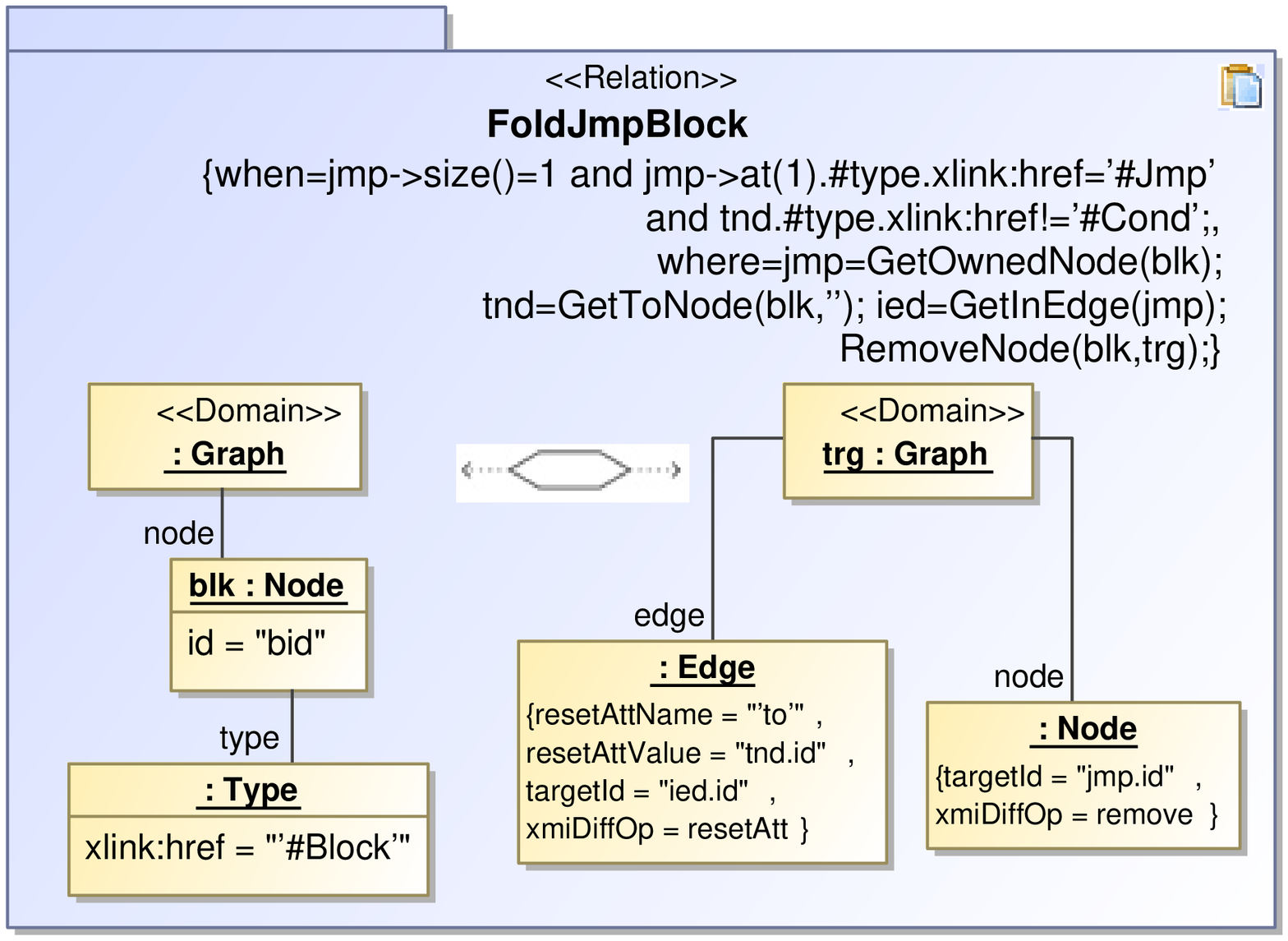}
\end{minipage}%
\\[+5pt]
\begin{minipage}[c]{0.4\linewidth}
\caption{Fold a \textsf{Const} without incoming edges (relName: \emph{FoldNode}, rInPlace : \emph{true})}
\label{fig:FoldIsolateConst}
\end{minipage}%
\hspace{6pt}
\begin{minipage}[c]{0.5\linewidth}
   \caption{Fold blocks containing only useless \textsf{Jmp} (relName: \emph{FoldNode}, rInPlace : \emph{true})}
\label{fig:FoldJmpBlock}
\end{minipage}
\end{figure}

\begin{figure}[!h]
\begin{minipage}[t]{0.4\linewidth}
\vspace{0pt} \centering
\includegraphics[width=1.0\linewidth]{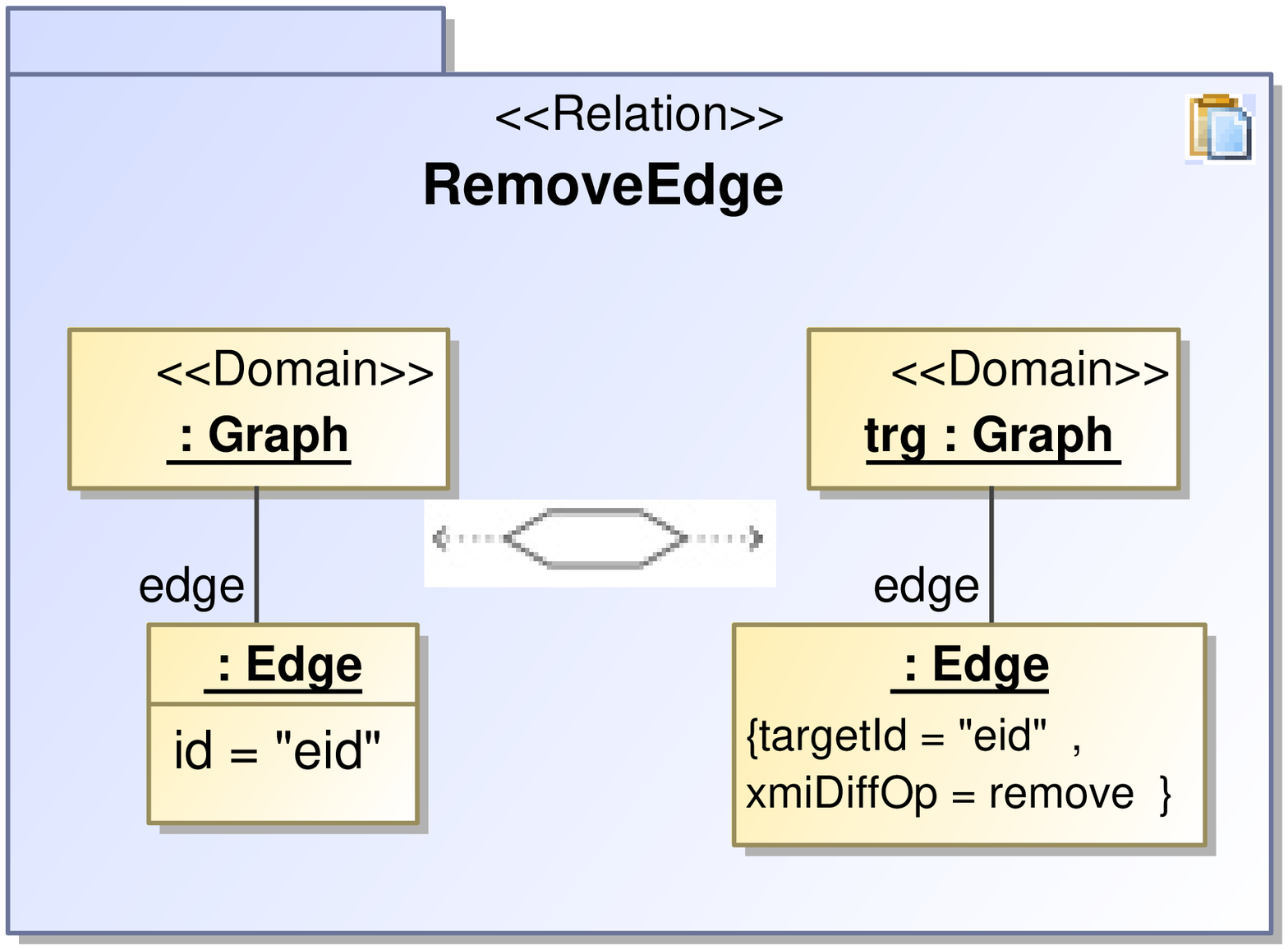}
\end{minipage}%
\begin{minipage}[t]{0.6\linewidth}
\vspace{0pt} \centering
\includegraphics[width=1.0\linewidth]{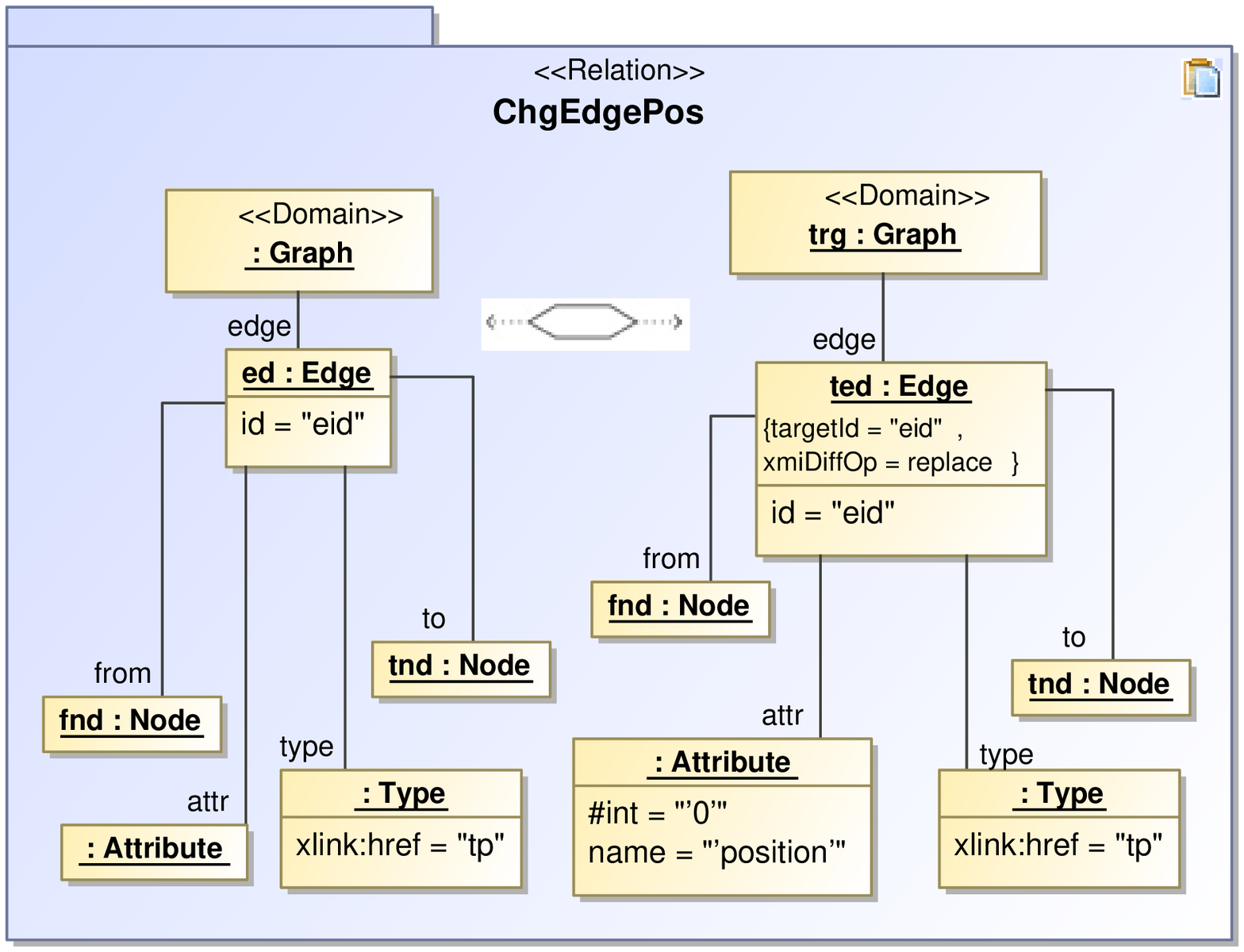}
\end{minipage}%
\\[+5pt]
\begin{minipage}[c]{0.4\linewidth}
\caption{Remove an \emph{Edge} }
\label{fig:RemoveEdge}
\end{minipage}%
\hspace{6pt}
\begin{minipage}[c]{0.5\linewidth}
   \caption{Change position of an \emph{Edge}}
\label{fig:ChgEdgePos}
\end{minipage}
\end{figure}

\begin{figure}[!h]
\begin{minipage}[t]{0.5\linewidth}
\vspace{0pt} \centering
\includegraphics[width=1.0\linewidth]{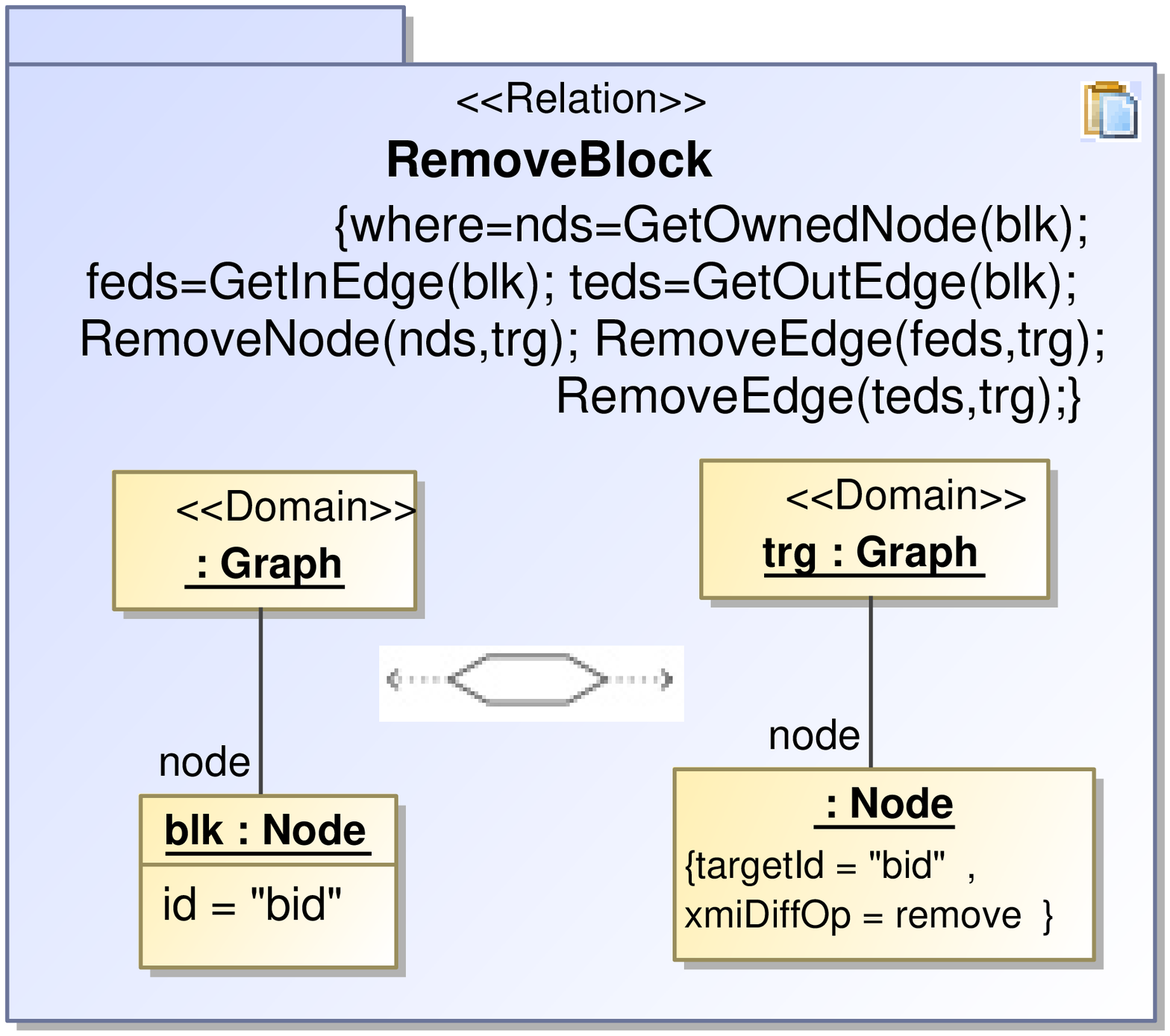}
\end{minipage}%
\begin{minipage}[t]{0.5\linewidth}
\vspace{0pt} \centering
\includegraphics[width=1.0\linewidth]{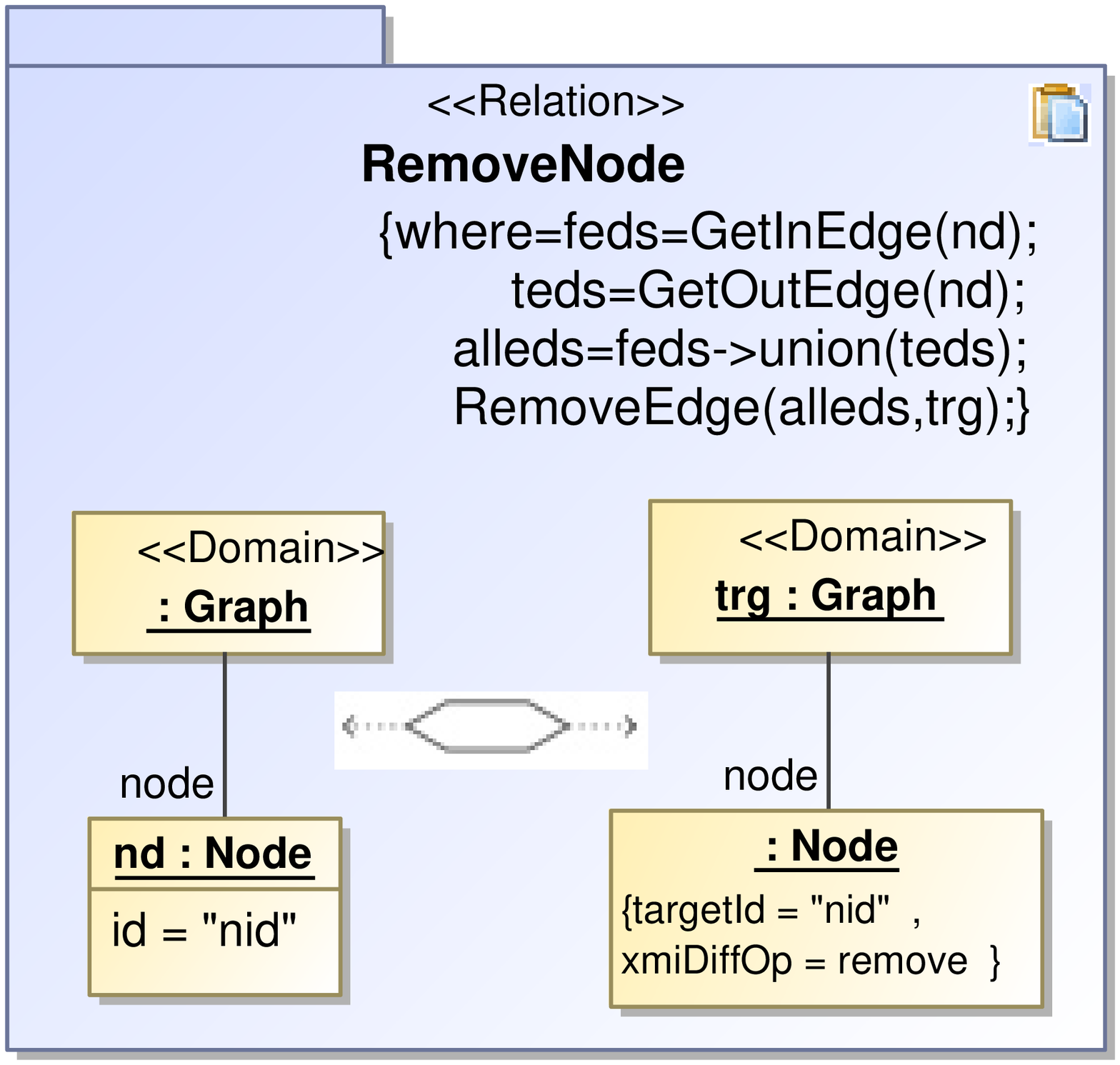}
\end{minipage}%
\\[+5pt]
\begin{minipage}[c]{0.4\linewidth}
\caption{Remove a \emph{Block} }
\label{fig:RemoveBlock}
\end{minipage}%
\hspace{6pt}
\begin{minipage}[c]{0.5\linewidth}
   \caption{Remove a \emph{Node}}
\label{fig:RemoveNode}
\end{minipage}
\end{figure}

\clearpage

\subsection{Queries}

\vspace*{-1.0\baselineskip}
\begin{figure}[!h]
\begin{minipage}[t]{0.5\linewidth}
\vspace{0pt} \centering
\includegraphics[width=.8\linewidth]{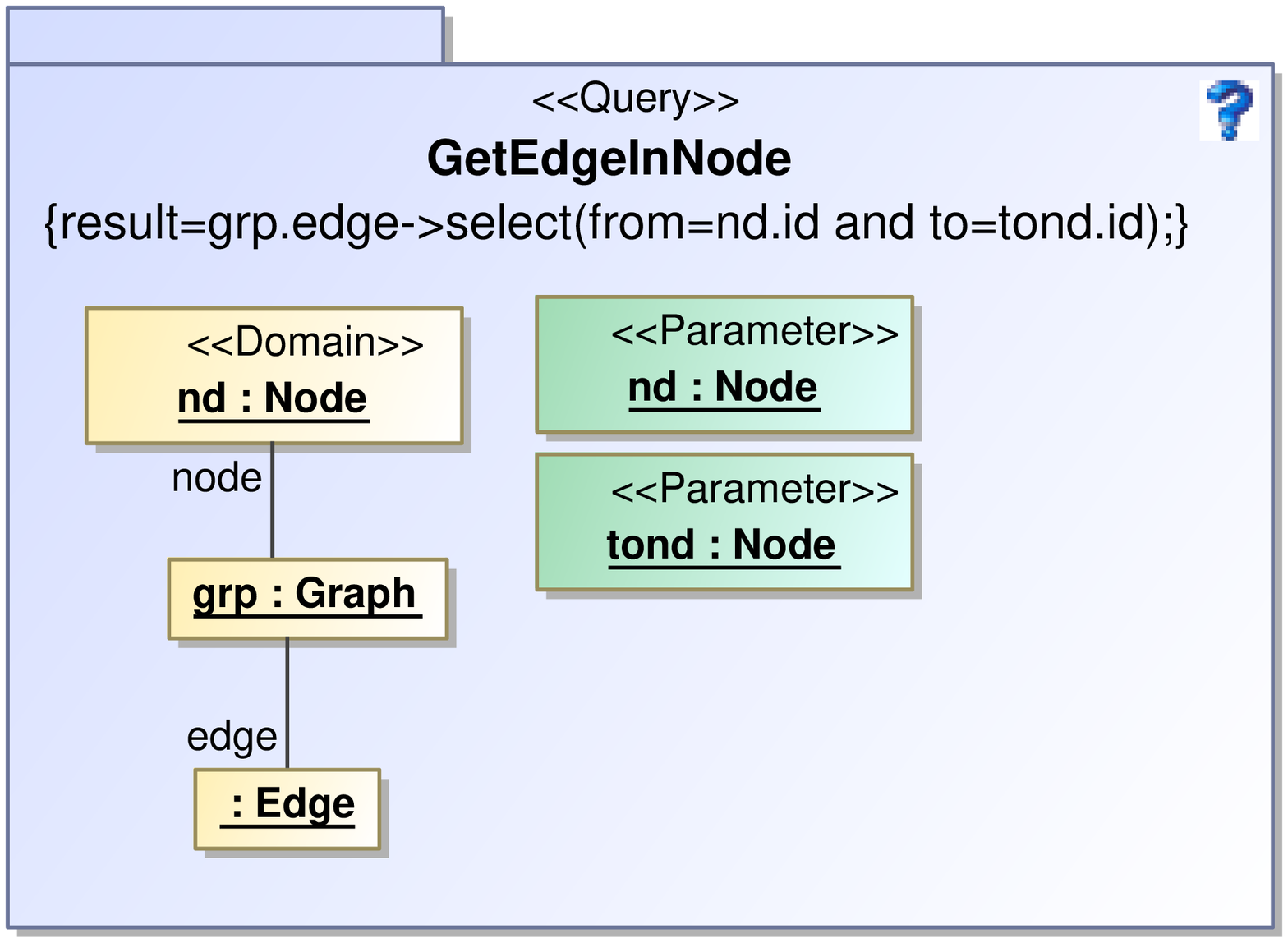}
\end{minipage}%
\begin{minipage}[t]{0.5\linewidth}
\vspace{0pt} \centering
\includegraphics[width=0.6\linewidth]{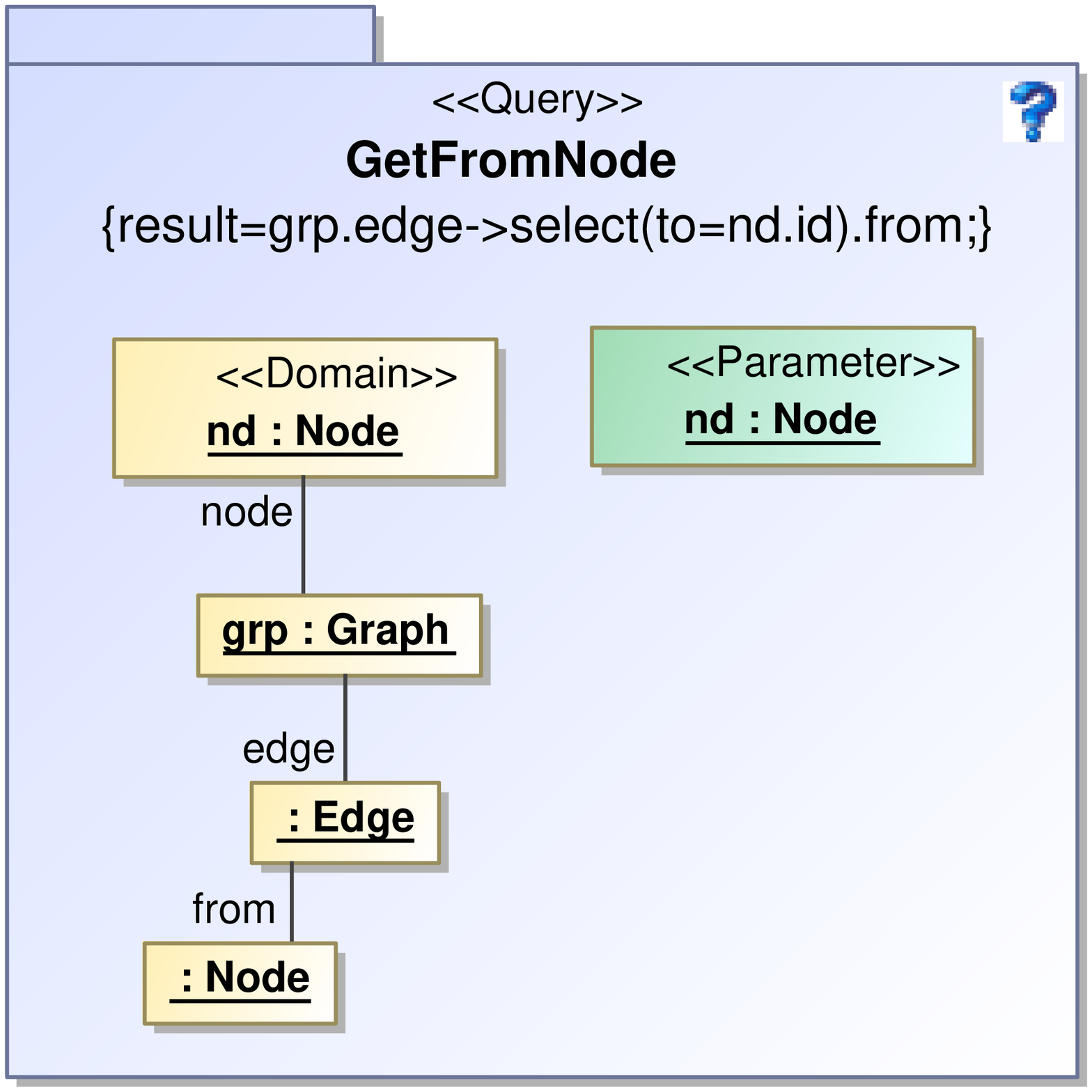}
\end{minipage}%
\\
\vspace*{-1.5\baselineskip}
\begin{minipage}[c]{0.4\linewidth}
\caption{Get edges between nodes}
\label{fig:GetEdgeInNode}
\end{minipage}%
\hspace{6pt}
\begin{minipage}[c]{0.5\linewidth}
   \caption{Get original nodes of a node}
\label{fig:GetFromNode}
\end{minipage}%
\\
\\
\begin{minipage}[t]{0.5\linewidth}
\vspace{0pt} \centering
\includegraphics[width=.8\linewidth]{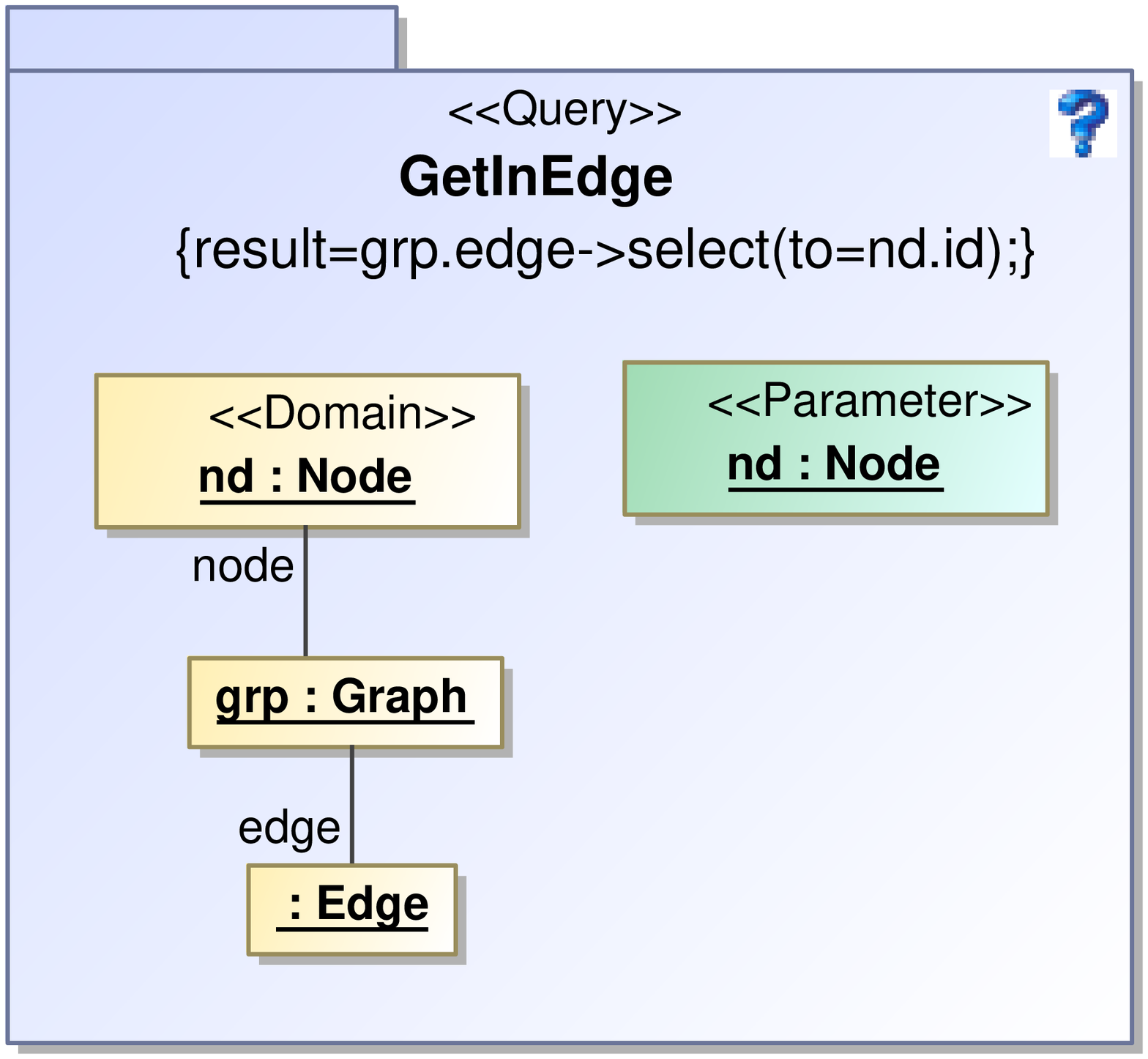}
\end{minipage}%
\begin{minipage}[t]{0.5\linewidth}
\vspace{0pt} \centering
\includegraphics[width=.8\linewidth]{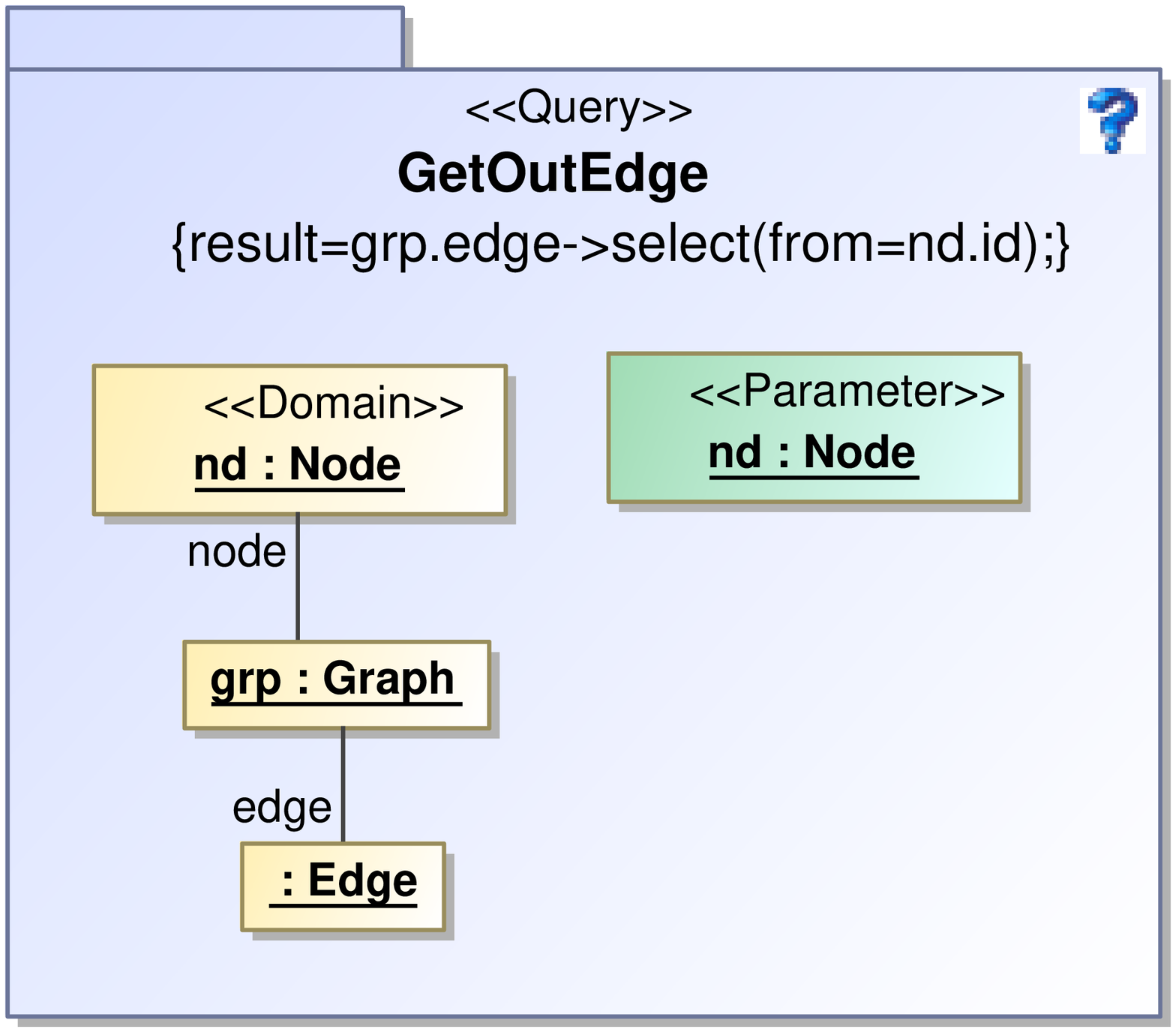}
\end{minipage}%
\\
\vspace*{-1.5\baselineskip}
\begin{minipage}[c]{0.4\linewidth}
\caption{Get incoming edges of a node}
\label{fig:GetInEdge}
\end{minipage}%
\hspace{6pt}
\begin{minipage}[c]{0.5\linewidth}
   \caption{Get outgoing edges of a node}
\label{fig:GetOutEdge}
\end{minipage}%
\\
\\
\\
\\
\begin{minipage}[t]{0.5\linewidth}
\vspace{0pt} \centering
\includegraphics[width=1.0\linewidth]{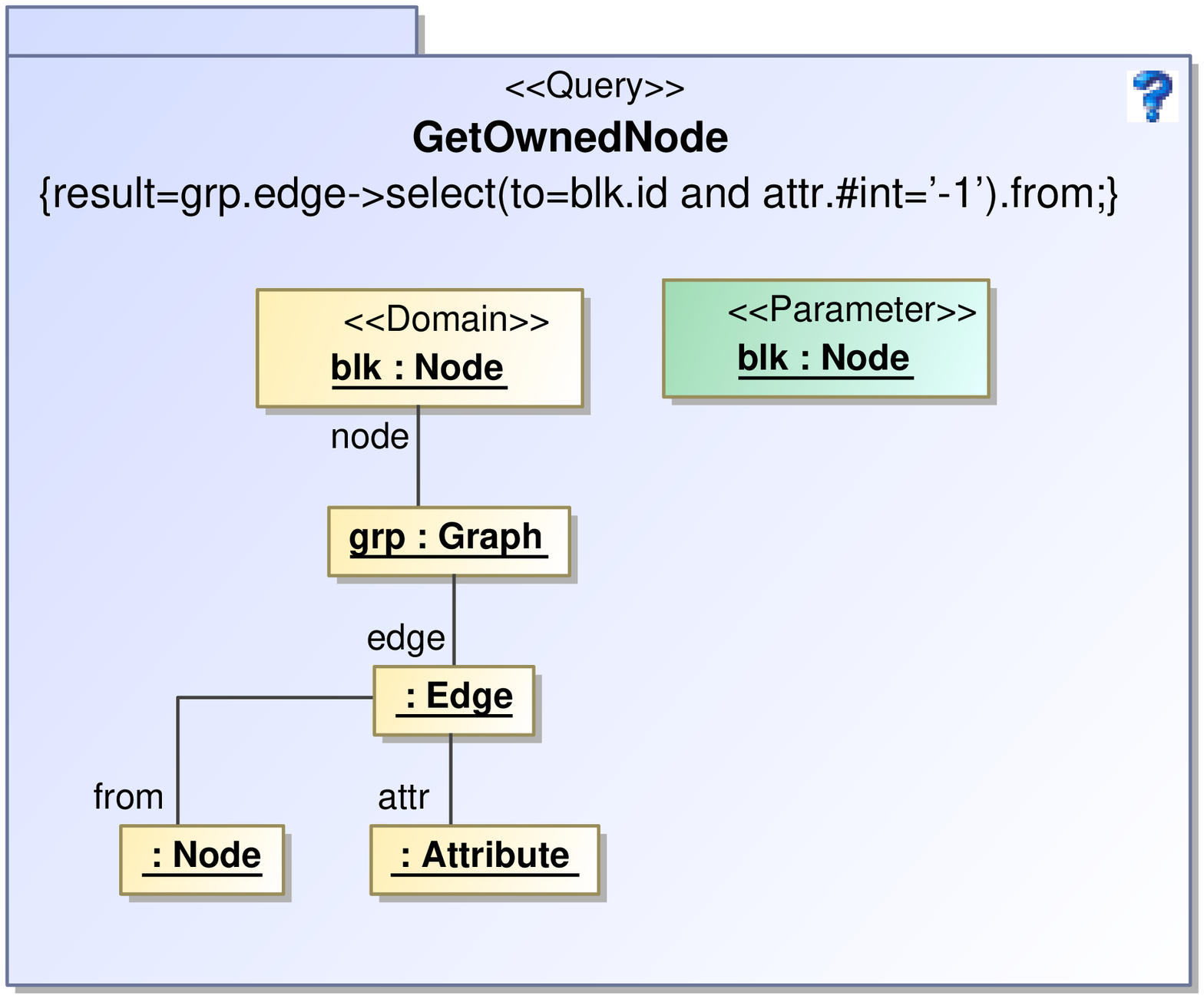}
\end{minipage}%
\begin{minipage}[t]{0.5\linewidth}
\vspace{0pt} \centering
\includegraphics[width=1.0\linewidth]{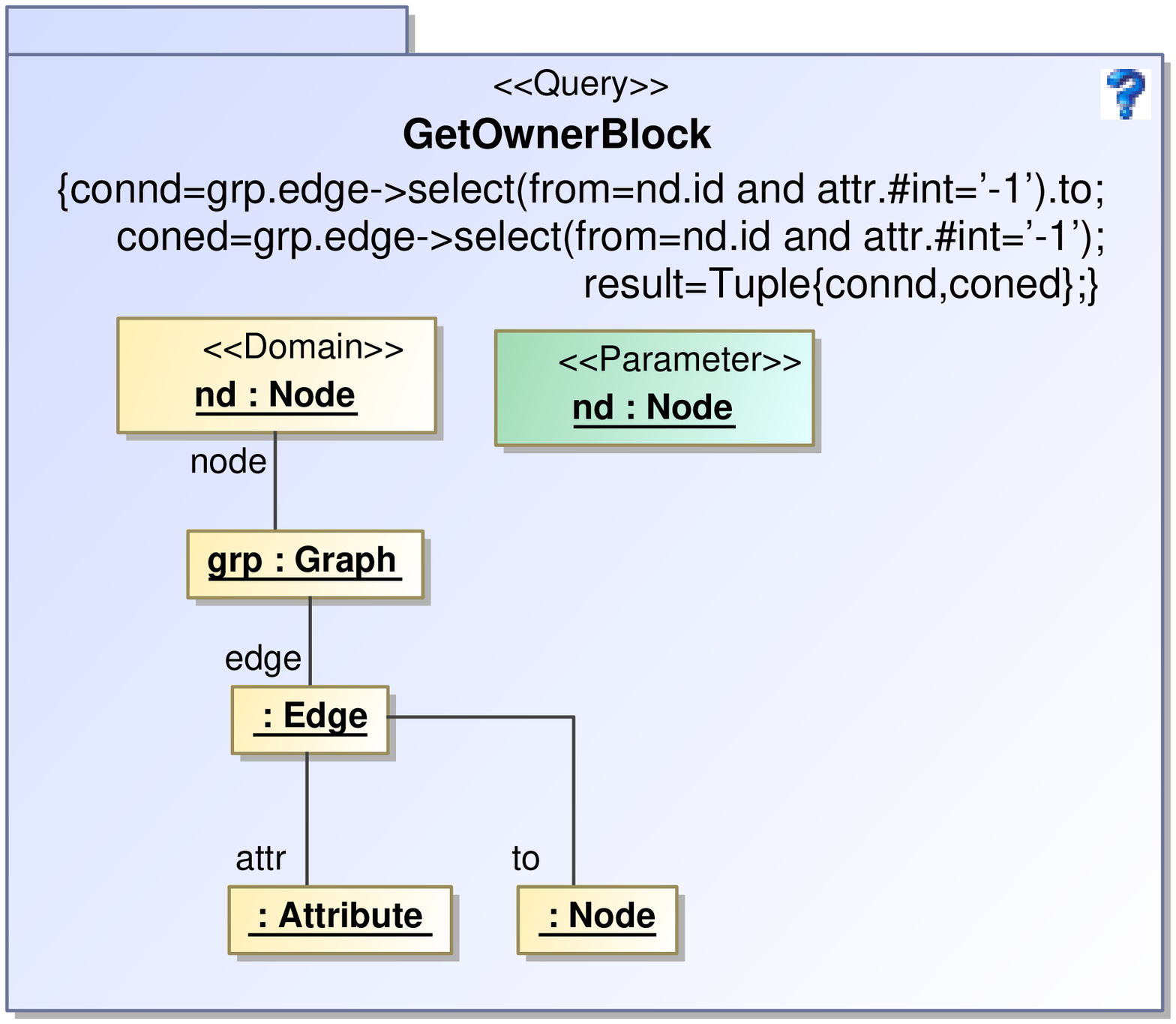}
\end{minipage}%
\\
\vspace*{-1.5\baselineskip}
\begin{minipage}[c]{0.4\linewidth}
\caption{Get owned nodes of a block}
\label{fig:GetOwnedNode}
\end{minipage}%
\hspace{6pt}
\begin{minipage}[c]{0.5\linewidth}
   \caption{Get owner block of a node}
\label{fig:GetOwnerBlock}
\end{minipage}%
\end{figure}

\begin{figure}[!t]
\begin{minipage}[t]{0.4\linewidth}
\vspace{0pt} \centering
\includegraphics[width=1.0\linewidth]{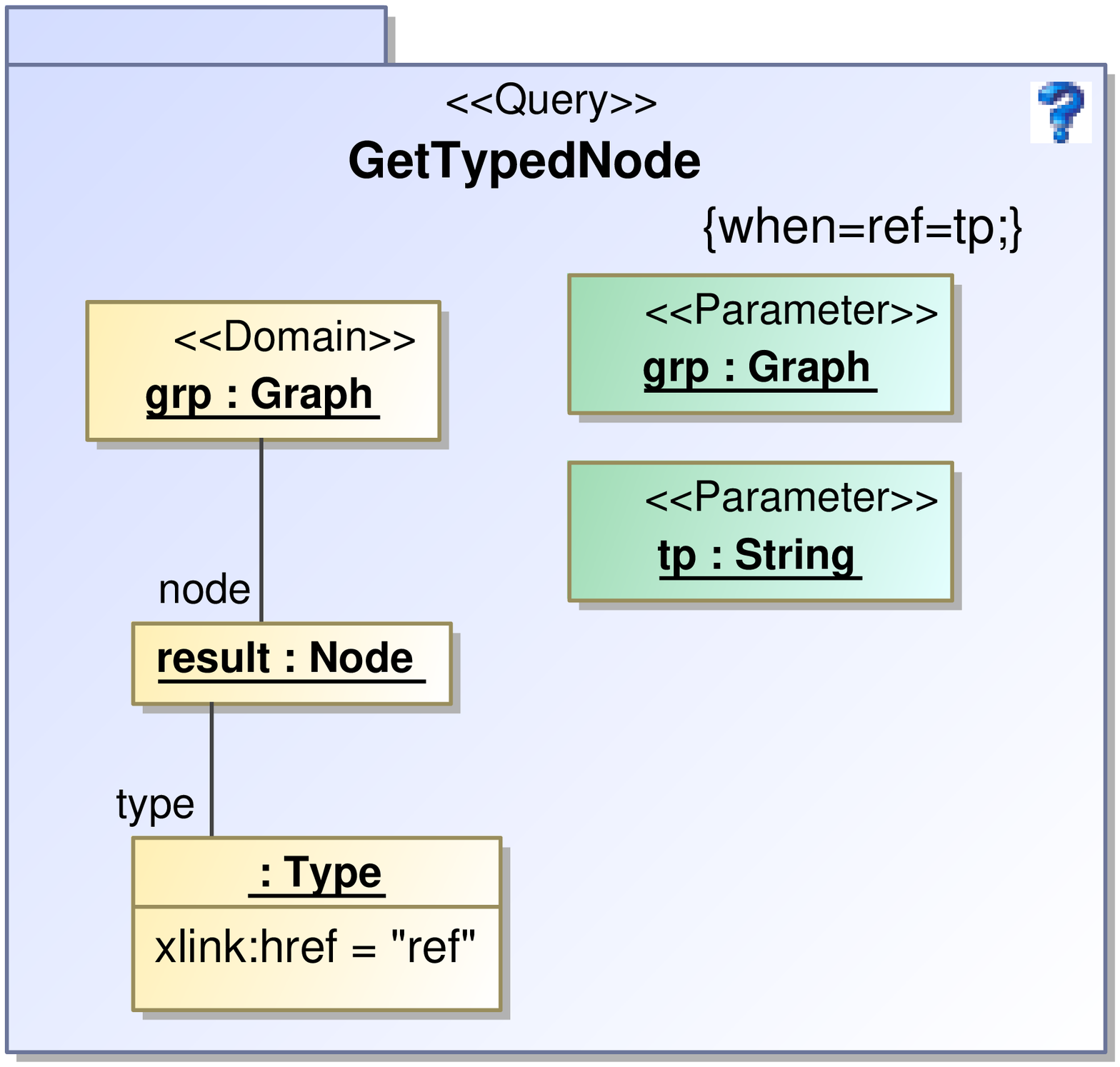}
\end{minipage}%
\begin{minipage}[t]{0.6\linewidth}
\vspace{0pt} \centering
\includegraphics[width=1.0\linewidth]{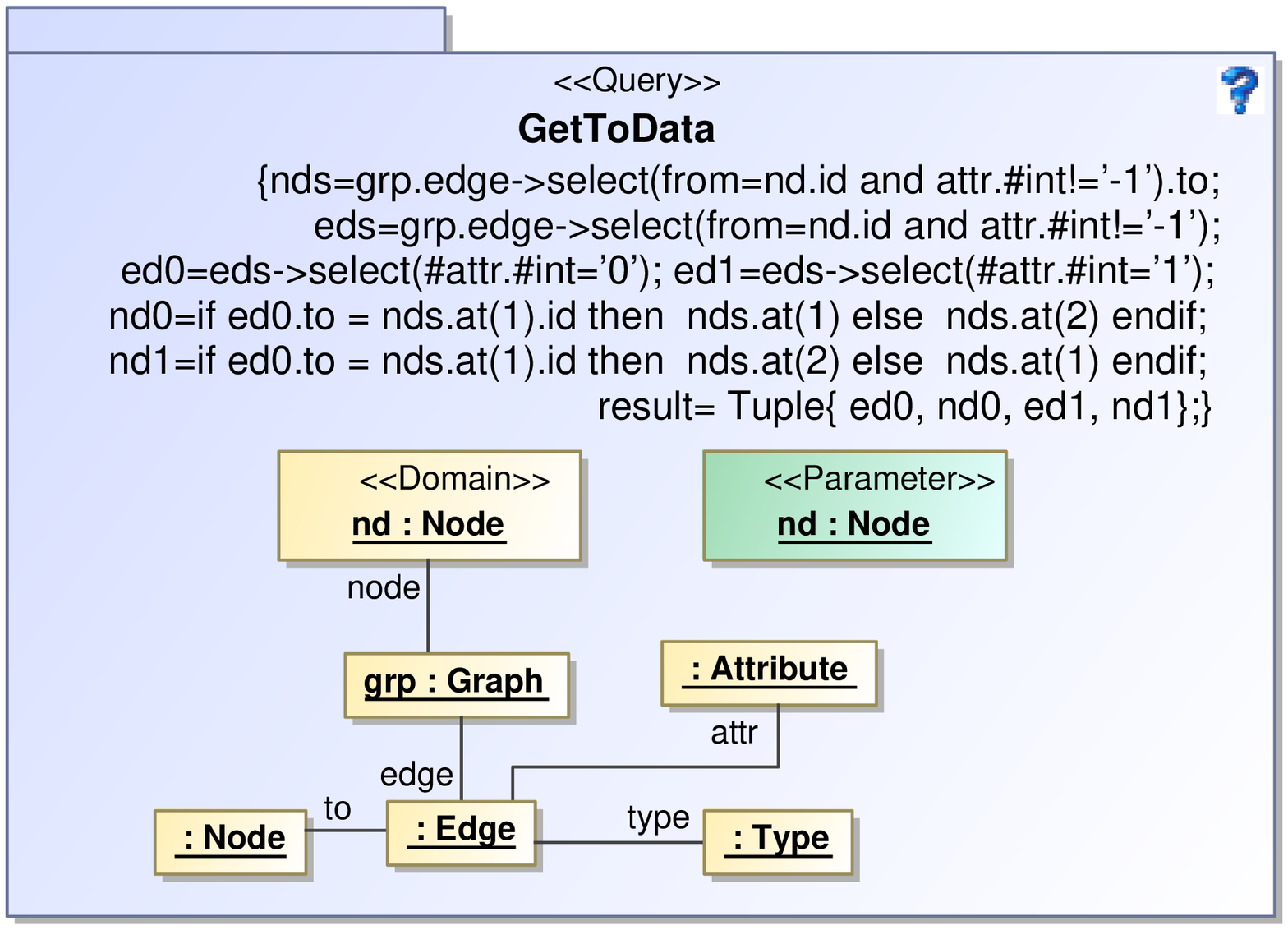}
\end{minipage}%
\\
\vspace*{-1.5\baselineskip}
\begin{minipage}[c]{0.4\linewidth}
\caption{Get nodes of specific type}
\label{fig:GetTypedNode}
\end{minipage}%
\hspace{6pt}
\begin{minipage}[c]{0.5\linewidth}
   \caption{Get edges and nodes of data operands}
\label{fig:GetToData}
\end{minipage}
\\
\\
\\
\begin{minipage}[t]{1.0\linewidth}
\vspace{0pt} \centering
\includegraphics[width=0.85\linewidth]{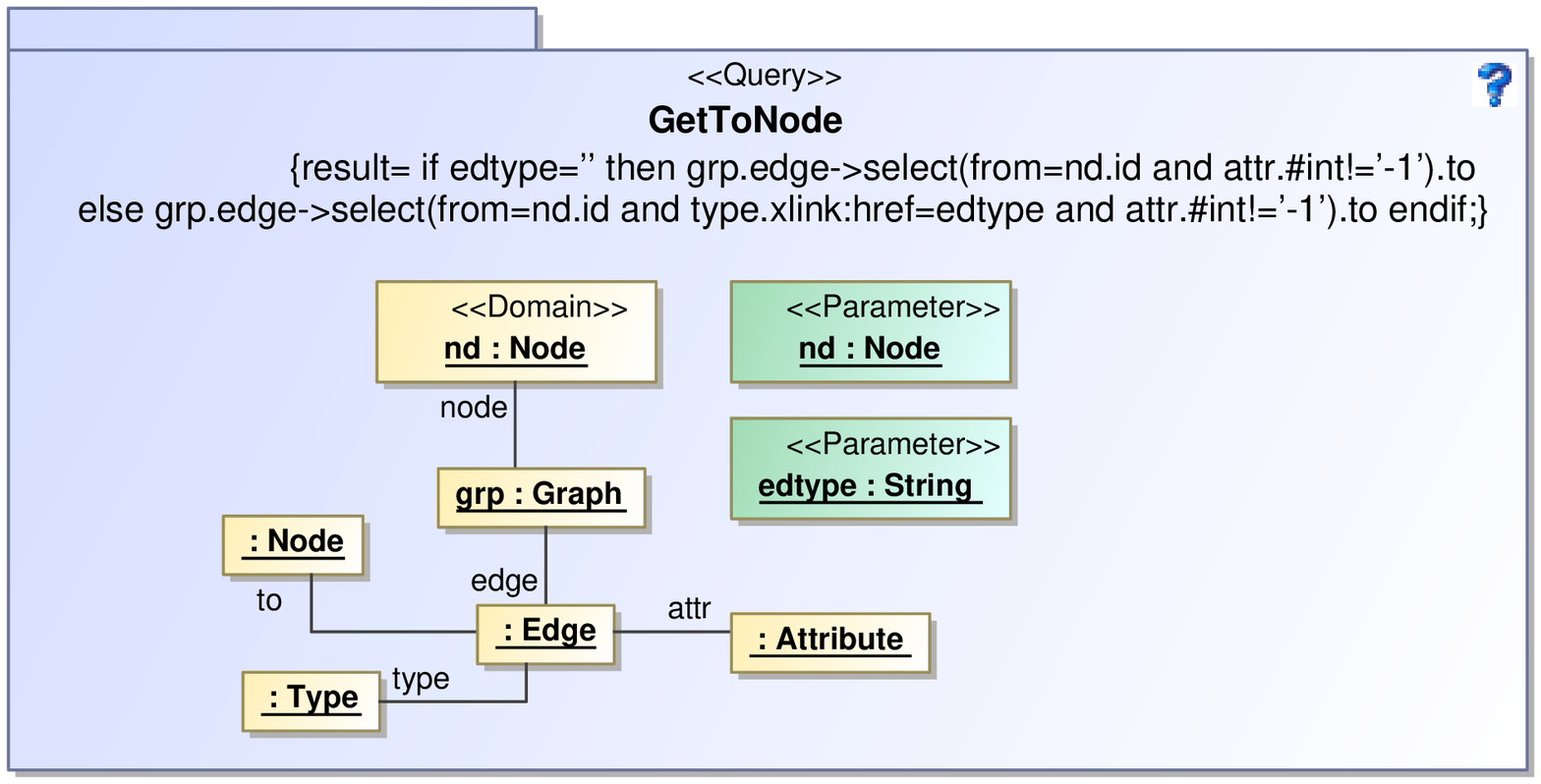}
\end{minipage}%
\\
\vspace*{-1.5\baselineskip}
\begin{minipage}[c]{1.0\linewidth}
\caption{Get destination nodes of a node}
\label{fig:GetToNode}
\end{minipage}%
\end{figure}
\clearpage

\subsection{Functions}

\begin{figure}[!h]
\begin{minipage}[c]{0.8\linewidth}
\begin{lstlisting}[language={},frame=single,numbers=none,basicstyle=\footnotesize,%
morekeywords={if, else, then, and, endif, result}, keywordstyle=\bfseries]

    less = if op='LESS' and v0 < v1 then 'true' else '' endif;
    noless = if op='LESS' and v0 > v1 then 'false' else '' endif;
    grt = if op='GREATER' and v0 > v1 then 'true' else '' endif;
    nogrt = if op='GREATER' and v0 < v1 then 'false' else '' endif;
    eq= if op='EQUAL' and v0 = v1 then 'true' else '' endif;
    noeq= if op='EQUAL' and v0 != v1 then 'false' else '' endif;

    result = less + noless + grt + nogrt + eq + noeq;
\end{lstlisting}
\caption{Function \textbf{CalcuLogic}(v0: Integer, v1: Integer, op:String) : String}
\label{fig:CalcuLogic}
\end{minipage}%
\\
\\
\\
\\
\begin{minipage}[c]{0.8\linewidth}
\begin{lstlisting}[language={},frame=single,numbers=none,basicstyle=\footnotesize,%
morekeywords={if, else, then, and, endif, result}, keywordstyle=\bfseries]

    add = if op='#Add' then v0 + v1 + 0 else 0 endif;
    sub = if op='#Sub' then v0 - v1  else 0 endif;
    mul = if op='#Mul' then v0 * v1  else 0 endif;
    div = if op='#Div' then v0 / v1  else 0 endif;

    result = add + sub + mul + div + 0;
\end{lstlisting}
\caption{Function \textbf{CalcuMatch}(v0: Integer, v1: Integer, op:String):Integer}
\label{fig:CalcuMatch}
\end{minipage}
\end{figure}

%% file: TTC2011_Appendix_sel.tex
\section{Transformation for instruction selection \label{ap:selecte}}

\noindent $\bullet$ \textbf{Transformation configuration:} name : \emph{TTC\_InstructionSelection}, source : \emph{Intermediate}, sourceKey : \emph{id}, sourceName : \emph{srcgrp}, target: \emph{Intermediate}, targetKey:\emph{id}, targetName : \emph{trggrp}.

\vspace*{-0.5\baselineskip}

\subsection{QVTR relations}

\vspace*{-1.5\baselineskip}

\begin{figure}[!h]
\begin{minipage}[t]{0.5\linewidth}
\vspace{0pt} \centering
\includegraphics[width=1.0\linewidth]{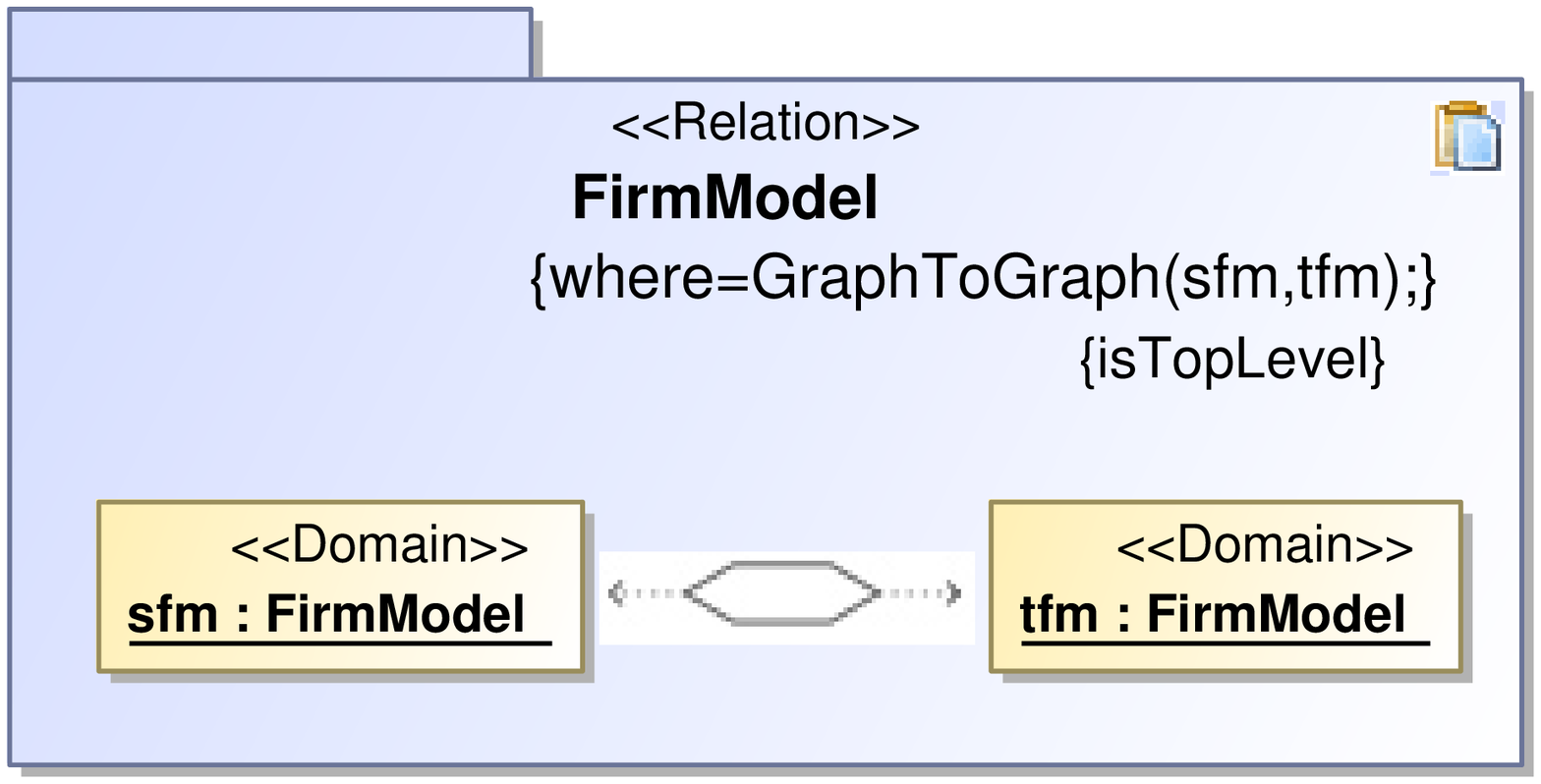}
\end{minipage}%
\begin{minipage}[t]{0.5\linewidth}
\vspace{0pt} \centering
\includegraphics[width=1.0\linewidth]{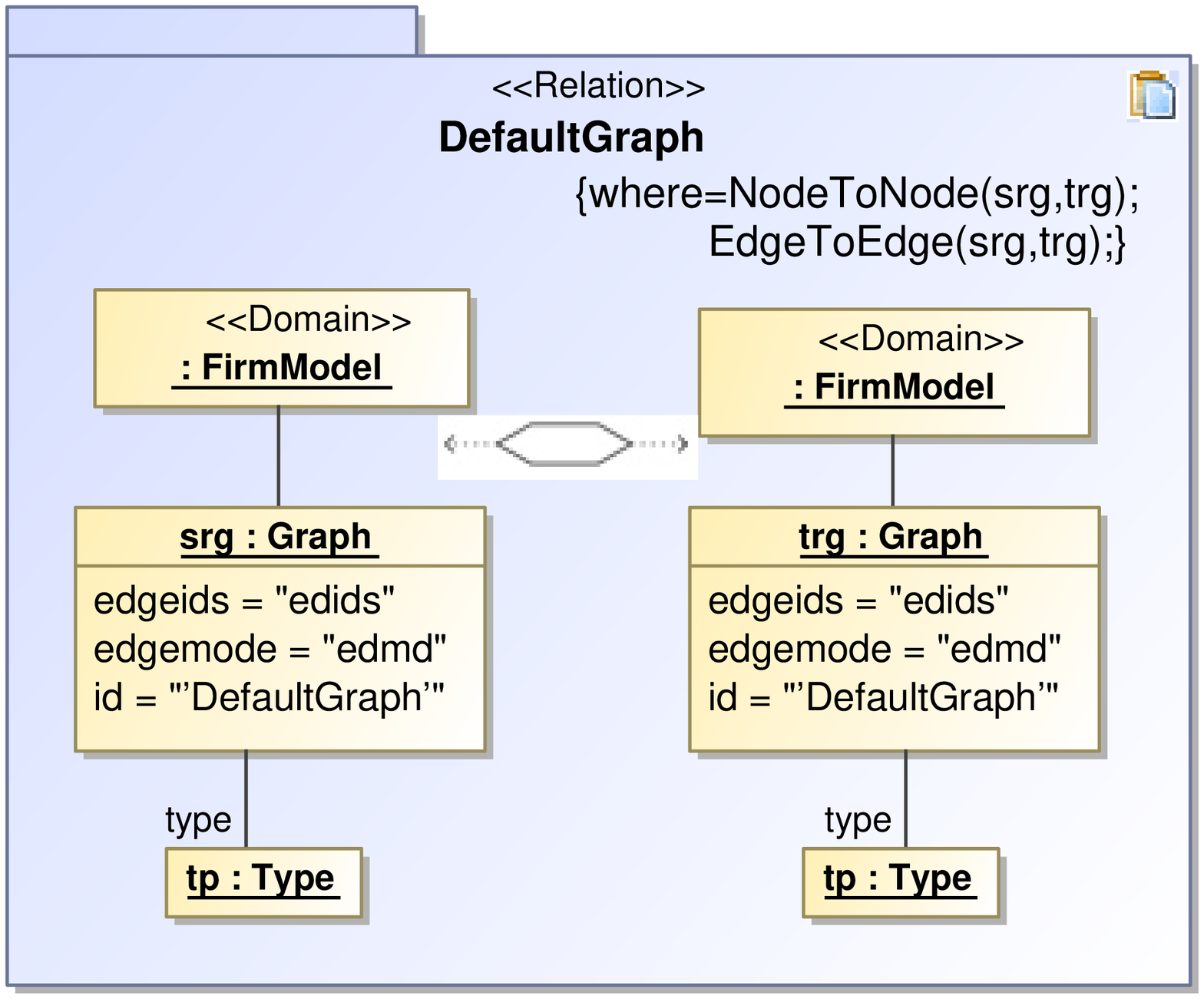}
\end{minipage}%
\\[+2pt]
\begin{minipage}[c]{0.5\linewidth}
\caption{Starting top level relation}
\label{fig:FirmModel}
\end{minipage}%
\begin{minipage}[c]{0.5\linewidth}
   \caption{Cope with default graph (relName : \emph{GraphToGraph})}
\label{fig:DefaultGraph}
\end{minipage}
\\[+20pt]
\begin{minipage}[t]{0.5\linewidth}
\vspace{0pt} \centering
\includegraphics[width=1.0\linewidth]{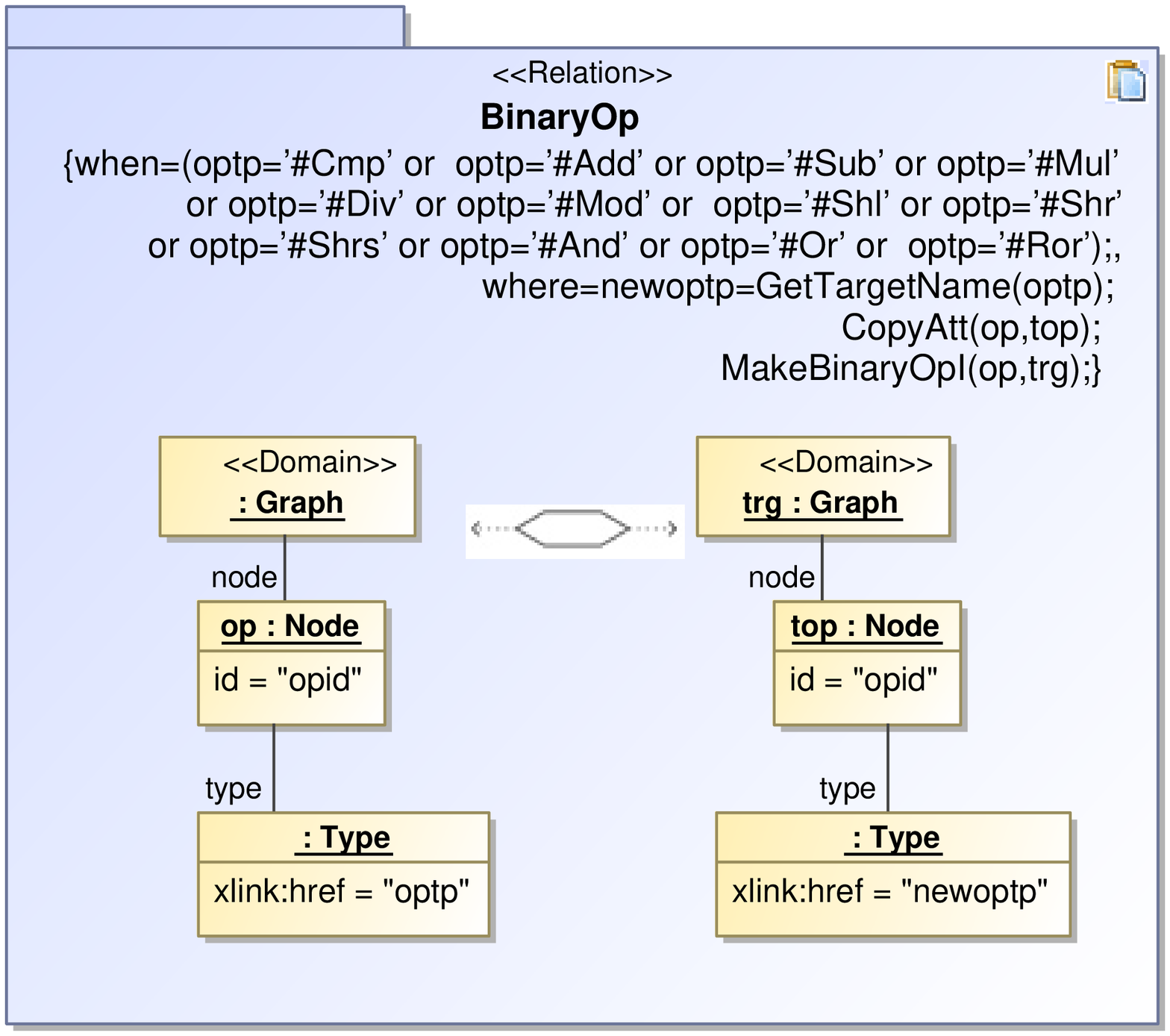}
\end{minipage}%
\begin{minipage}[t]{0.5\linewidth}
\vspace{0pt} \centering
\includegraphics[width=1.0\linewidth]{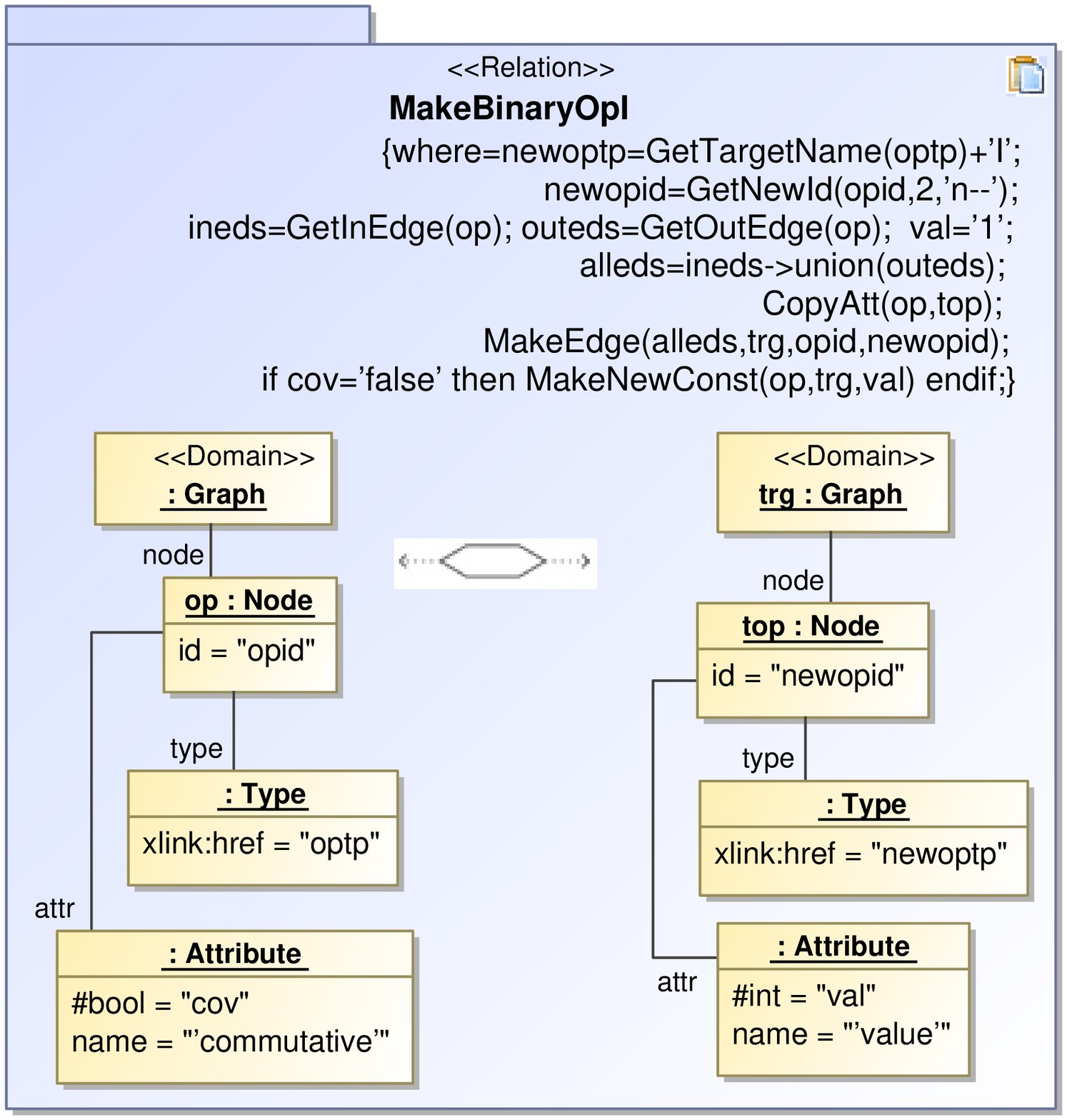}
\end{minipage}%
\\[+5pt]
\begin{minipage}[c]{0.4\linewidth}
\caption{Select and cope with binary operations (relName: \emph{NodeToNode})}
\label{fig:BinaryOp}
\end{minipage}%
\hspace{6pt}
\begin{minipage}[c]{0.5\linewidth}
   \caption{Create the \textsf{TargetOpI} node}
\label{fig:MakeBinaryOpI}
\end{minipage}
\end{figure}

\begin{figure}[!h]
\begin{minipage}[t]{0.4\linewidth}
\vspace{0pt} \centering
\includegraphics[width=1.0\linewidth]{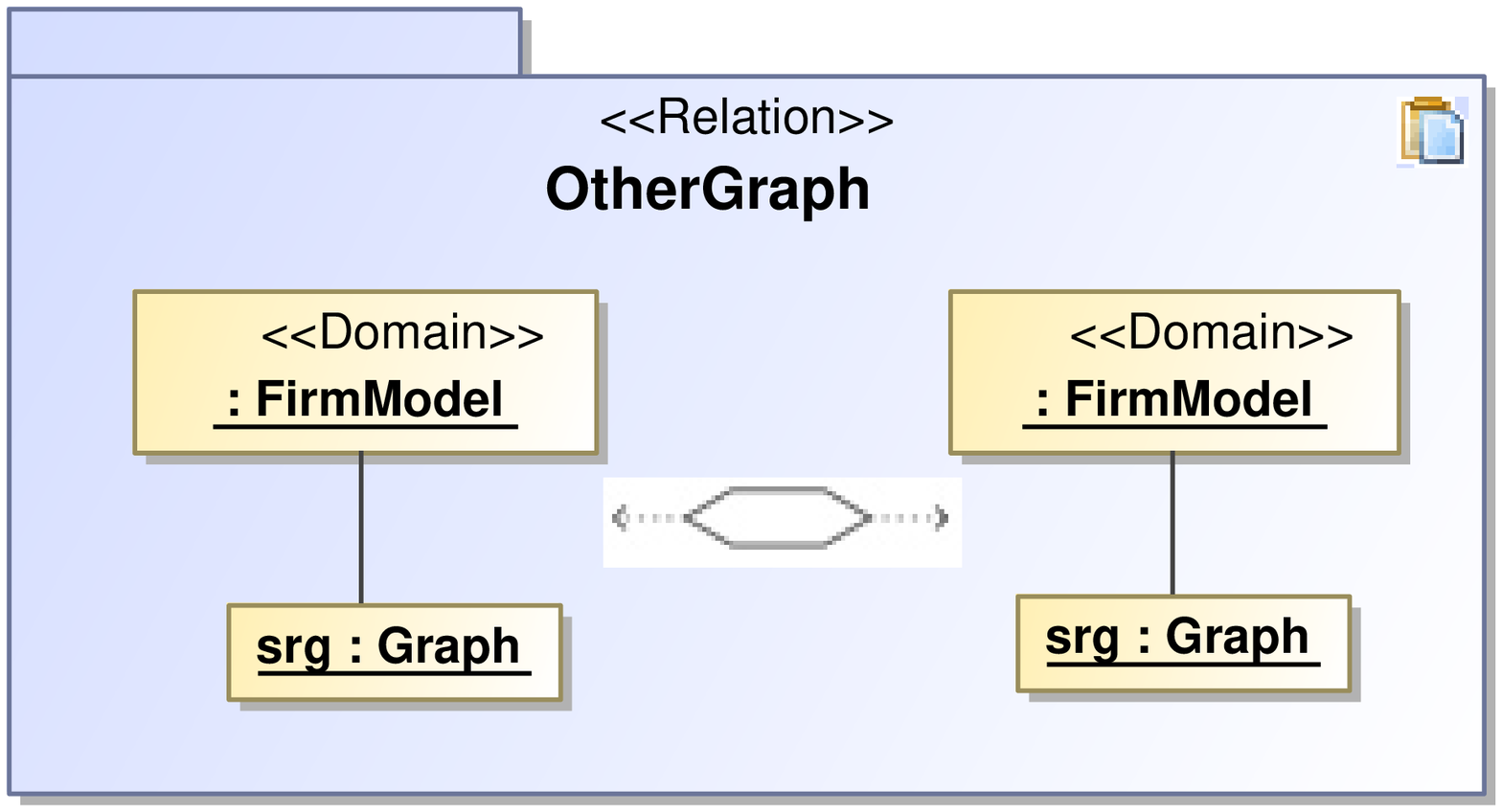}
\end{minipage}%
\begin{minipage}[t]{0.6\linewidth}
\vspace{0pt} \centering
\includegraphics[width=.9\linewidth]{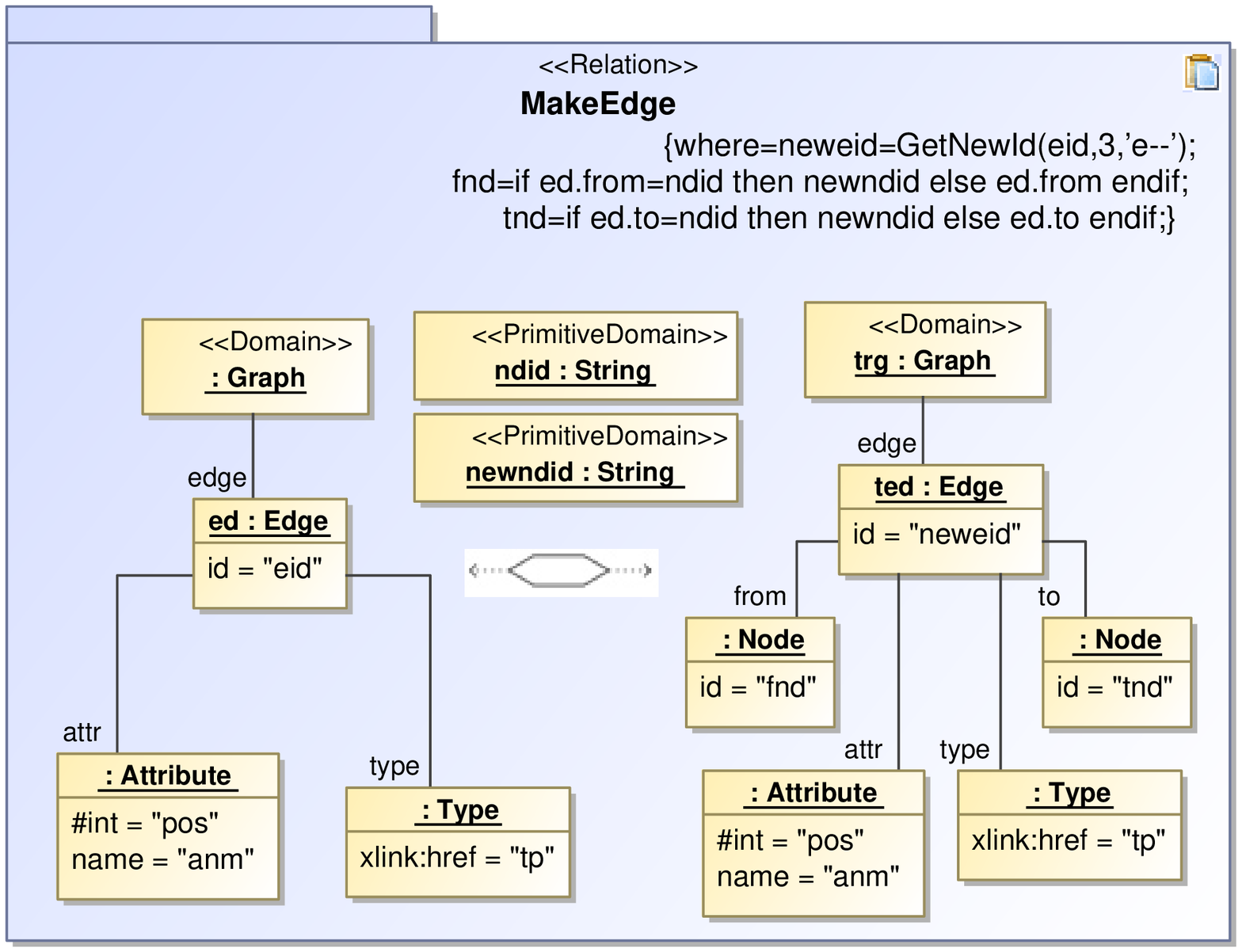}
\end{minipage}%
\\
\begin{minipage}[c]{0.4\linewidth}
\caption{Copy graphs other than the default one (relName: \emph{GraphToGraph})}
\label{fig:OtherGraph}
\end{minipage}%
\hspace{6pt}
\begin{minipage}[c]{0.5\linewidth}
   \caption{Create edges for the \textsf{TargetOpI} node}
\label{fig:MakeEdge}
\end{minipage}
\\[+15pt]
\begin{minipage}[t]{0.4\linewidth}
\vspace{0pt} \centering
\includegraphics[width=.9\linewidth]{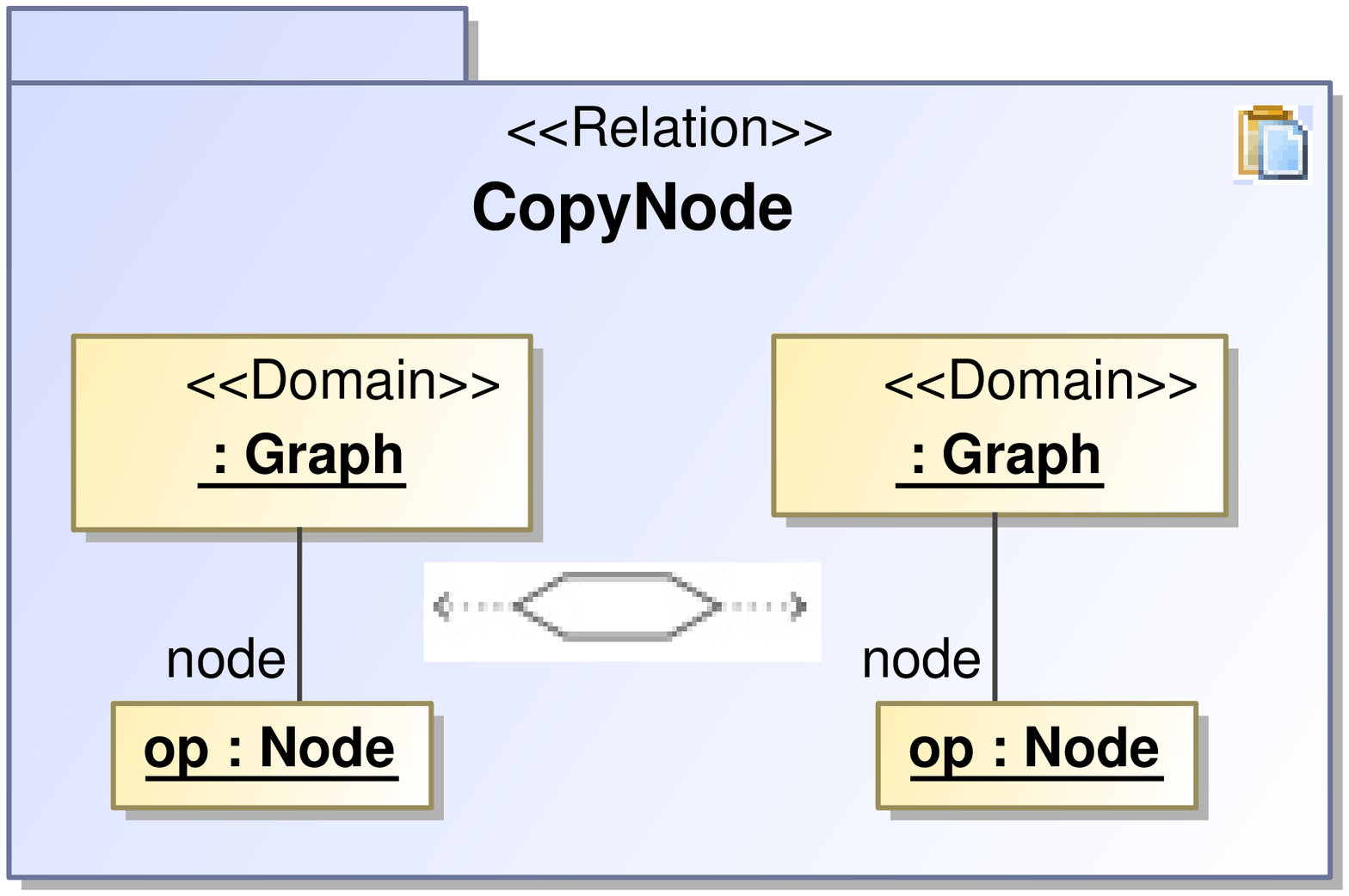}
\end{minipage}%
\begin{minipage}[t]{0.6\linewidth}
\vspace{0pt} \centering
\includegraphics[width=1.0\linewidth]{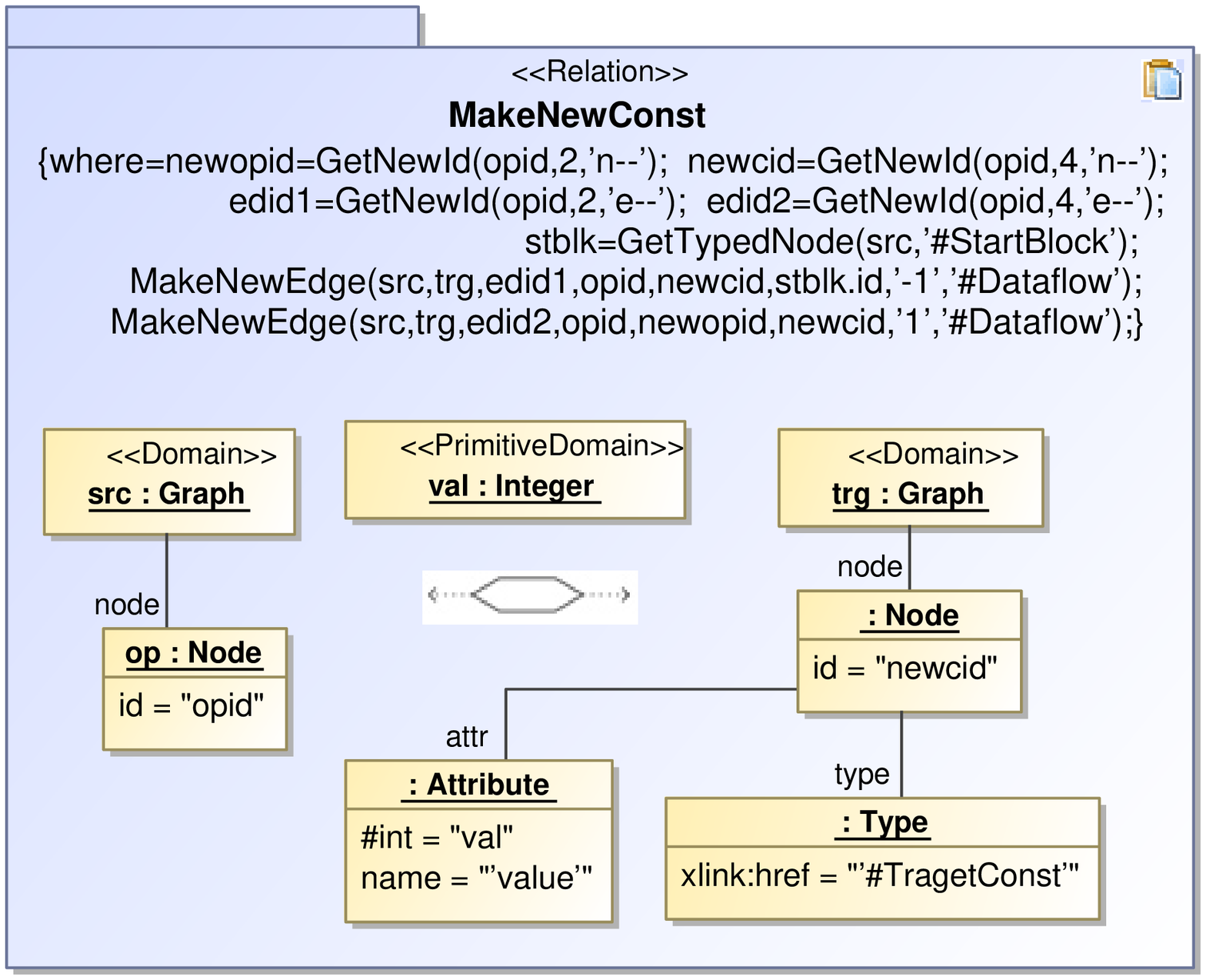}
\end{minipage}%
\\
\begin{minipage}[c]{0.4\linewidth}
\caption{Copy node of other type (relName: \emph{NodeToNode}) }
\label{fig:CopyNode}
\end{minipage}%
\hspace{6pt}
\begin{minipage}[c]{0.5\linewidth}
   \caption{Create a new const node for the \textsf{TargetOpI} node}
\label{fig:MakeNewConst}
\end{minipage}
\\[+15pt]
\begin{minipage}[t]{0.4\linewidth}
\vspace{0pt} \centering
\includegraphics[width=1.0\linewidth]{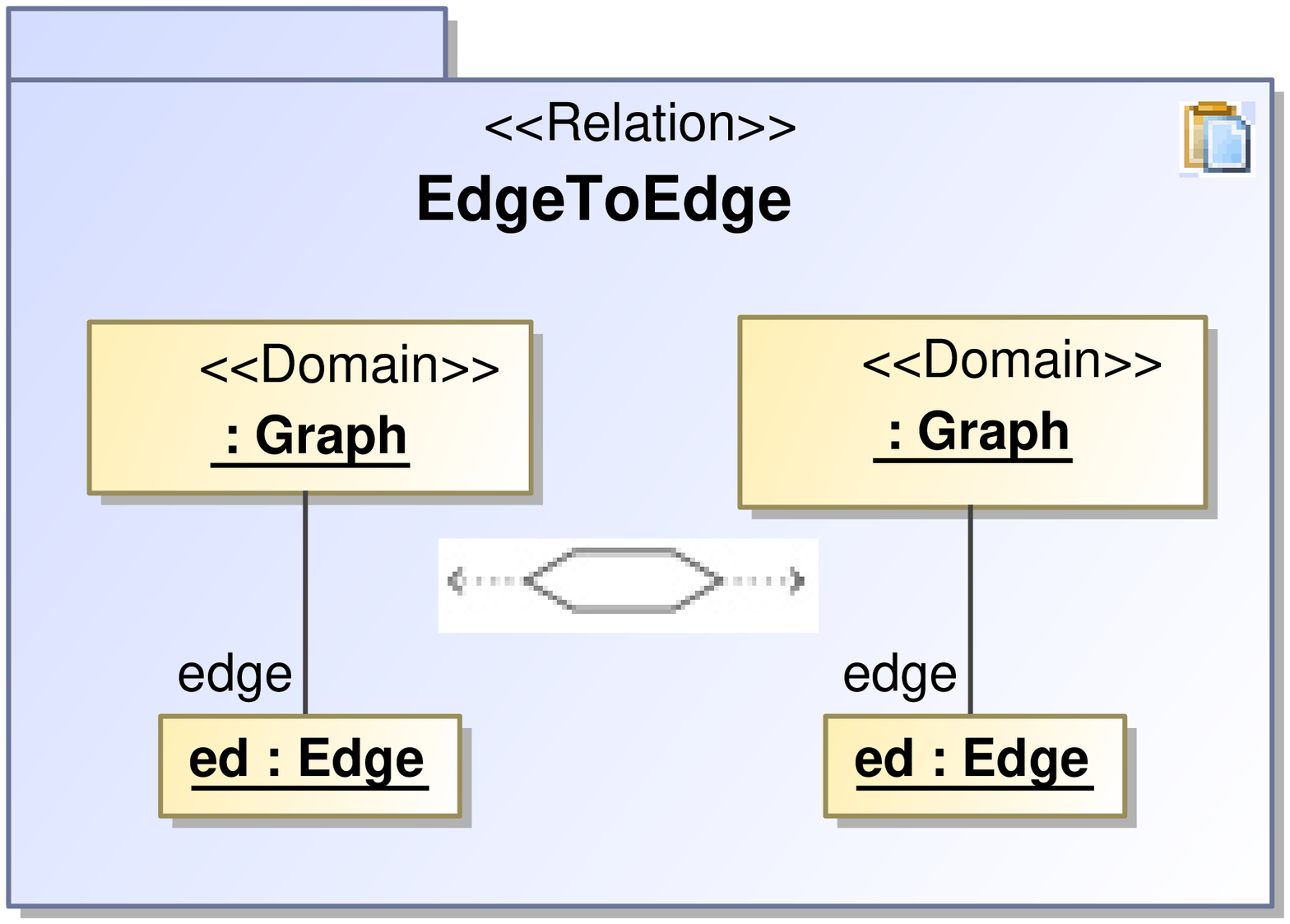}
\end{minipage}%
\begin{minipage}[t]{0.6\linewidth}
\vspace{0pt} \centering
\includegraphics[width=.9\linewidth]{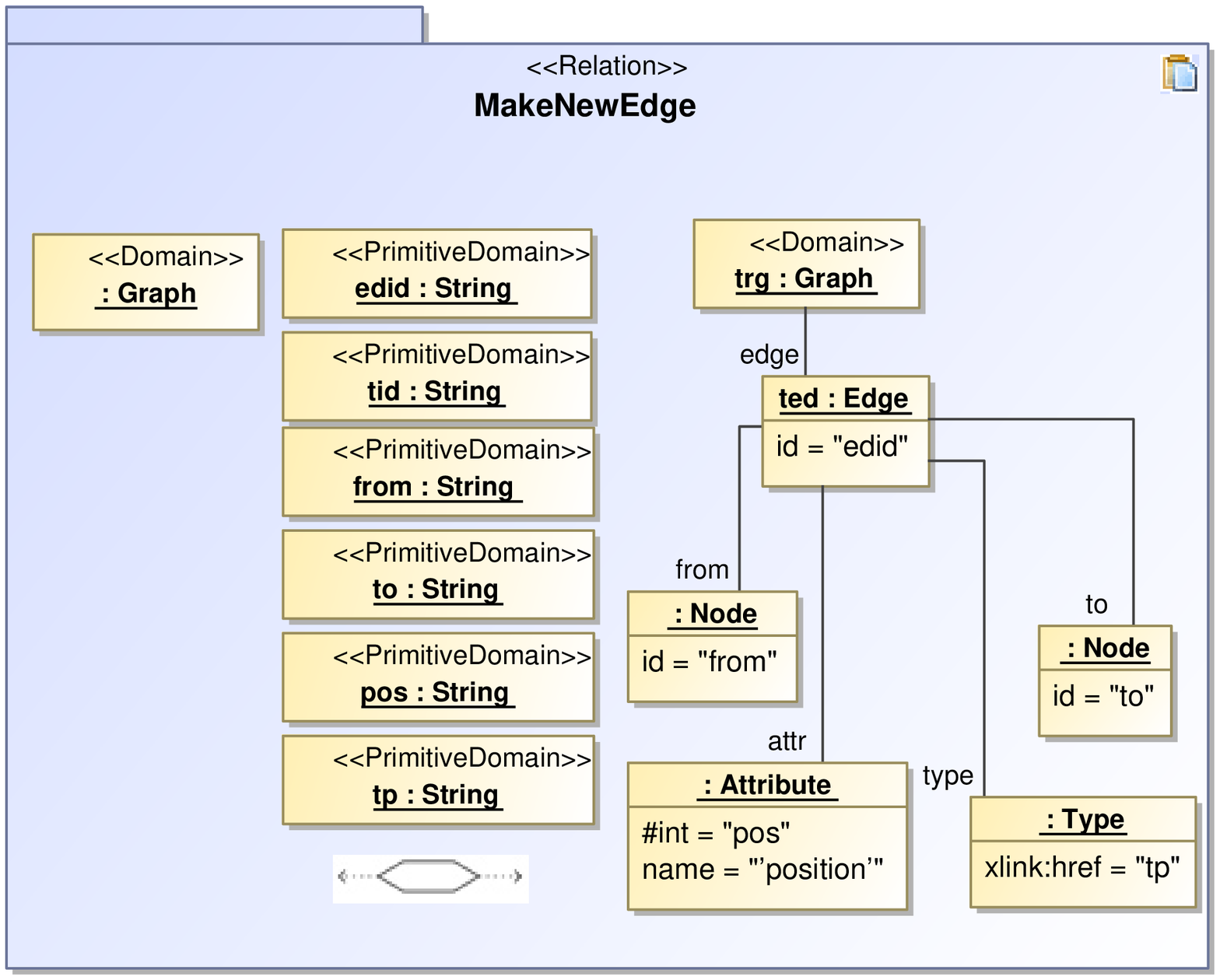}
\end{minipage}%
\\
\begin{minipage}[c]{0.4\linewidth}
\caption{Copy an edge}
\label{fig:EdgeToEdge}
\end{minipage}%
\hspace{6pt}
\begin{minipage}[c]{0.5\linewidth}
   \caption{Create a new edge}
\label{fig:MakeNewEdge}
\end{minipage}
\end{figure}
\clearpage

\begin{figure}[!h]
\begin{minipage}[t]{0.4\linewidth}
\vspace{0pt} \centering
\includegraphics[width=1.0\linewidth]{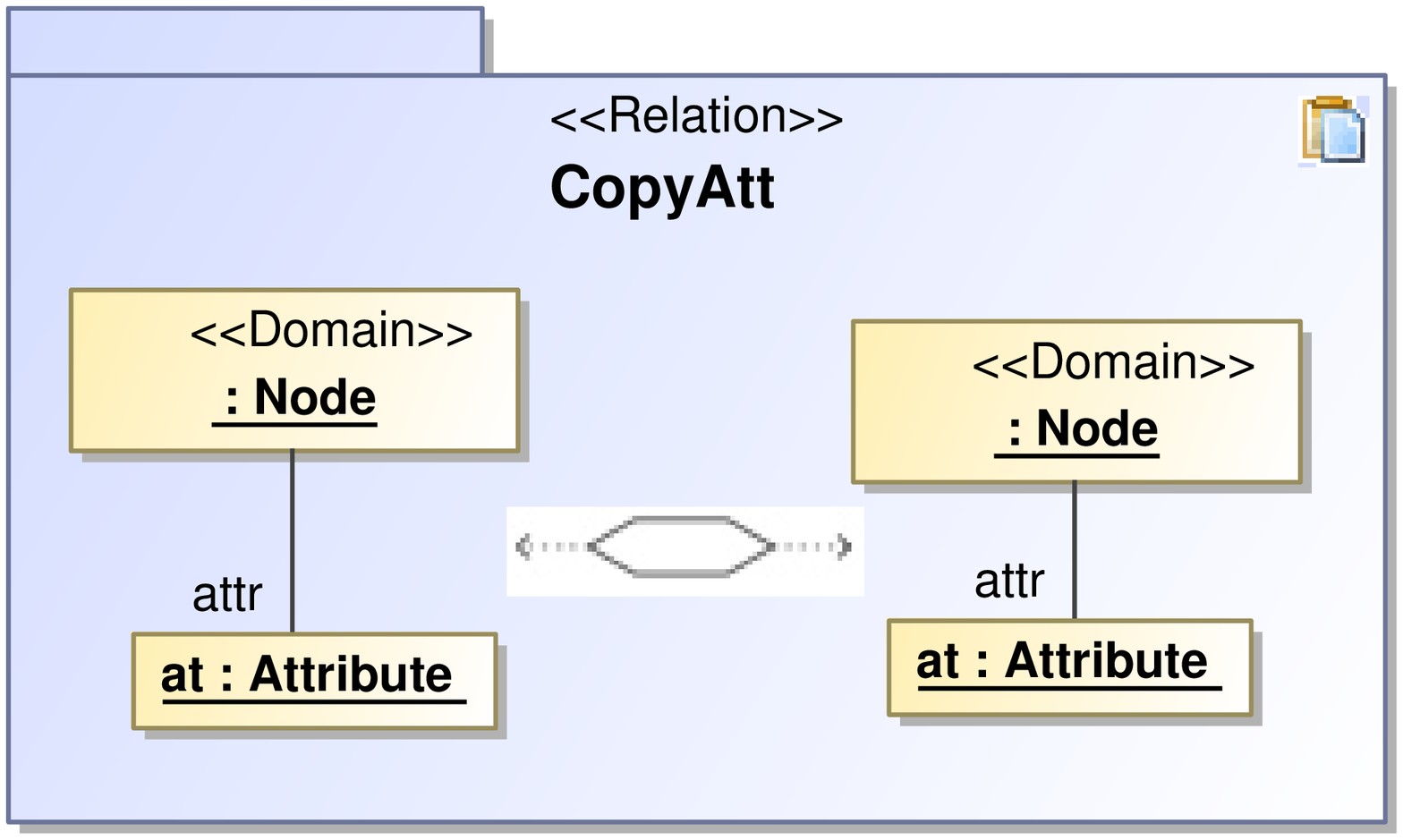}
\end{minipage}%
\begin{minipage}[t]{0.6\linewidth}
\vspace{0pt} \centering
\includegraphics[width=.9\linewidth]{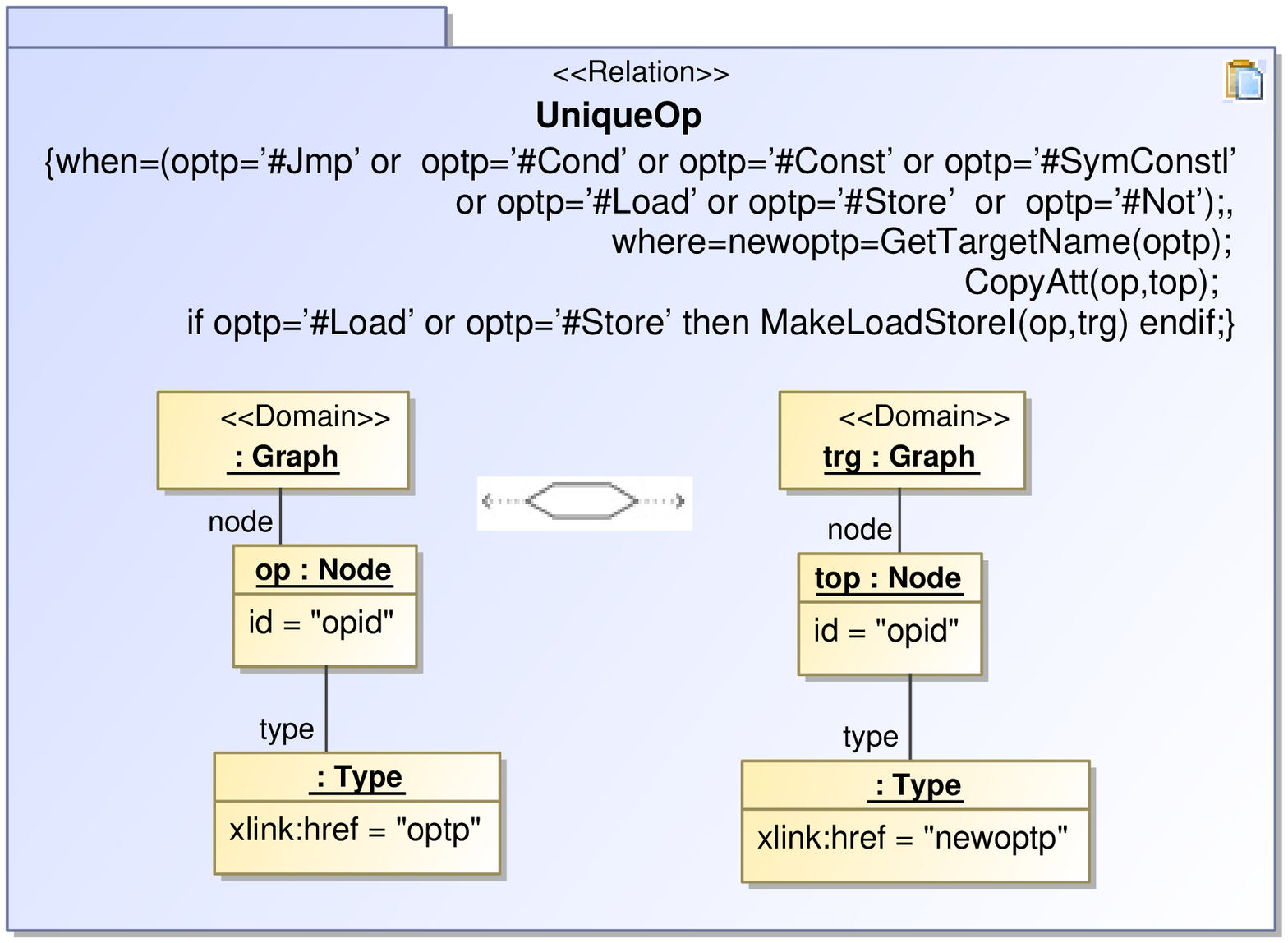}
\end{minipage}%
\\
\vspace*{-1.5\baselineskip}
\begin{minipage}[c]{0.4\linewidth}
\caption{Copy attribute of node}
\label{fig:CopyAtt}
\end{minipage}%
\hspace{6pt}
\begin{minipage}[c]{0.5\linewidth}
   \caption{Select and cope with other operations (relName: \emph{NodeToNode})}
\label{fig:UniqueOp}
\end{minipage}
\\[+20pt]
\begin{minipage}[t]{1.0\linewidth}
\vspace{0pt} \centering
\includegraphics[width=0.4\linewidth]{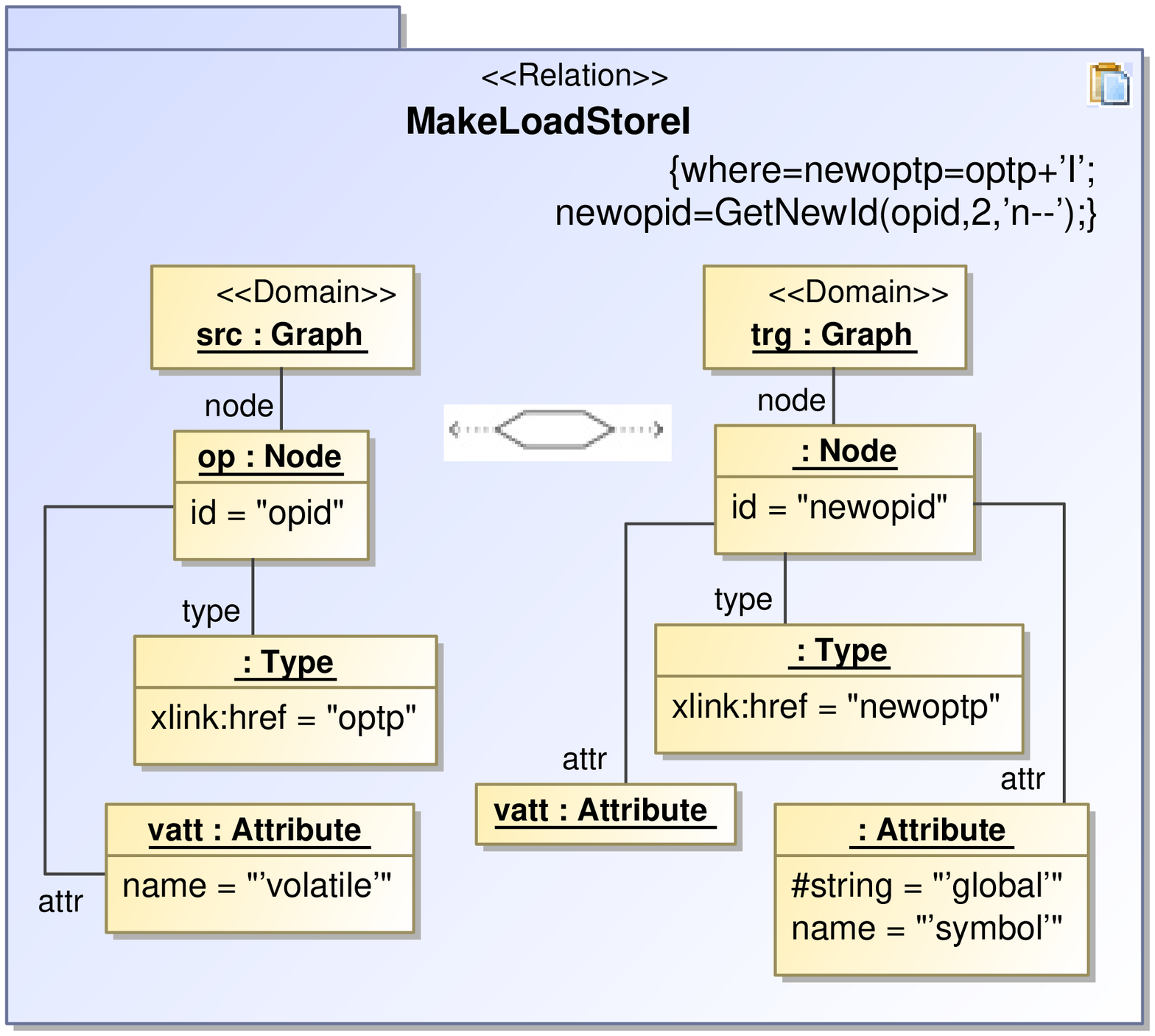}
\end{minipage}%
\\
\vspace*{-1.5\baselineskip}
\begin{minipage}[c]{1.0\linewidth}
\caption{Create \textsf{LoadI} or \textsf{StoreI} node}
\label{fig:MakeLoadStoreI}
\end{minipage}%
\end{figure}

\subsection{Queries and Functions}

All queries have same definitions as in transformation local optimizations:

$\bullet$ \textbf{GetInEdge} (Fig.~\ref{fig:GetInEdge}), \textbf{GetOutEdge} (Fig.~\ref{fig:GetOutEdge}), \textbf{GetOwnerBlock} (Fig.~\ref{fig:GetOwnerBlock})

$\bullet$ \textbf{GetToData} (Fig.~\ref{fig:GetToData}), \textbf{GetTypedNode} (Fig.~\ref{fig:GetTypedNode})

\begin{figure}[!h]
\begin{minipage}[c]{0.4\linewidth}
\begin{lstlisting}[language={},frame=single,numbers=none,basicstyle=\footnotesize,%
morekeywords={if, else, then, and, endif, result}, keywordstyle=\bfseries]
    pos1=pos+1;
    p1=substring(in,1,pos);
    p2=substring(in,pos1);
    result=sufix+p2+p1;
\end{lstlisting}
\vspace*{-1.0\baselineskip}
\caption{Function \textbf{GetNewId}(in: String, pos: Integer, sufix:String)}
\label{label=fig:GetNewId}
\end{minipage}%
\hspace{16pt}
\begin{minipage}[c]{0.4\linewidth}
\begin{lstlisting}[language={},frame=single, numbers=none,basicstyle=\footnotesize,%
morekeywords={if, else, then, and, endif, result}, keywordstyle=\bfseries]

    nm=substring-after(op,'#');
    result='#'+'Target'+nm;

\end{lstlisting}
\vspace*{-1.0\baselineskip}
\caption{Function \textbf{GetTargetName}(op : String)}
\label{fig:GetTargetName}
\end{minipage}
\end{figure}